\documentclass[12pt]{article}

\usepackage{graphicx}

\usepackage{booktabs}  %
\usepackage{xcolor}

\usepackage[colorlinks,citecolor=blue]{hyperref}
\usepackage{cite}
\usepackage{amsmath,amsfonts,amssymb}
\usepackage{float}
\usepackage{latexsym}
\usepackage{enumitem}
\usepackage{bm}

\usepackage{wrapfig}

\usepackage{amsthm}
\usepackage{xspace}

\usepackage{cleveref}
\usepackage{array}
\usepackage[normalem]{ulem}
\usepackage{comment}
\usepackage{bbm}

\usepackage{scalerel}
\usepackage{mathtools}
\usepackage{cases}
\usepackage{relsize}

\input{sstt_arxiv_sty.sty}

\newcommand{\uvector}{\mathbf{e}}

\newcommand{\kOtheta}{k_{0\theta}}

\newcommand{\zd}{\z_\text{d}}
\newcommand{\td}{t_\text{d}}

\newcommand{\xid}{\xi_\text{d}}
\newcommand{\zetad}{\zeta_\text{d}}
\newcommand{\etad}{\eta_\text{d}}
\newcommand{\zdO}{z_{\text{d}1}}
\newcommand{\zdT}{z_{\text{d}2}}
\newcommand{\nuz}{\nu_\text{b}}
\newcommand{\nut}{\nu_t}
\newcommand{\nus}{\nu_s}
\newcommand{\nuinst}{\nu_\text{inst}}
\newcommand{\muz}{\mu_\text{b}}
\newcommand{\mut}{\mu_t}
\newcommand{\mus}{\mu_s}
\newcommand{\zetamax}{\zeta_{\max}}
\newcommand{\zetamin}{\zeta_{\min}}
\newcommand{\zetay}{\zeta_\y}
\newcommand{\zetaym}{\zeta_{\y,m}}
\newcommand{\xis}{\xi_s}

\newcommand{\sigmasq}{\sigma^2}
\newcommand{\sigmasqz}{\sigma^2_\text{b}}
\newcommand{\sigmasqt}{\sigma^2_t}

\newcommand{\sigmasqs}{\sigma^2_s}
\newcommand{\sigmasqalpha}{\sigma^2_\alpha}
\newcommand{\sigmasqn}{\sigma^2_\text{n}}

\newcommand{\IT}{I^T}
\newcommand{\ITalpha}{I^T_\alpha}
\newcommand{\ITz}{I^T_\text{b}}

\newcommand{\IS}{I^S}
\newcommand{\ISalpha}{I^S_\alpha}
\newcommand{\ISz}{I^S_\text{b}}

\newcommand{\phiT}{\varphi_T}

\newcommand{\tmax}{{t_{\max}}}

\newcommand{\streak}{\text{streak}}
\newcommand{\Nimg}{N_\text{img}}

\newcommand{\avg}[1]{\left\langle#1\right\rangle}

\newcommand{\Tlinphi}{T^\varphi_\text{lin}}
\newcommand{\Wlin}{W_\text{lin}}
\newcommand{\Ilin}{I_\text{lin}}
\newcommand{\bPhi}{b_\Phi}
\newcommand{\una}{\text{U}}

\newcommand{\Phiarg}[2]{\Phi\big(\kOtheta\phiT #1, \kOtheta\phiT^2 #2\big)}
\newcommand{\abssqPhiarg}[2]{\big|\Phi\big(\kOtheta\phiT #1, \kOtheta\phiT^2 #2 \big)\big|^2}

\newcommand{\roverline}[1]{\mathpalette\doroverline{#1}}
\newcommand{\doroverline}[2]{\overline{#1#2}}

\DeclareMathOperator{\sign}{sign}
\DeclareMathOperator{\std}{std}

\DeclareMathAlphabet{\brm}{OT1}{cmr}{bx}{rm}

\newcommand\mapsfrom{\mathrel{\reflectbox{\ensuremath{\mapsto}}}}

\begin{document}

\title{Detection of delayed target response in SAR}

\author{Mikhail Gilman and Semyon Tsynkov 
\\ 
Department of Mathematics, 
\\
North Carolina State University, 
\\
Campus Box 8205, Raleigh, NC 27695, USA}

\maketitle

\begin{abstract}
Delayed target response in synthetic aperture radar (SAR) imaging can be obscured by the range-delay ambiguity and speckle. To analyze the range-delay ambiguity, one
extends the standard SAR formulation and allows both the target reflectivity and the image to depend not only on the coordinates, but also on the response delay. However, this still leaves the speckle unaccounted for.   
Yet speckle is  commonly found in SAR images of extended targets, and a statistical approach is usually employed to describe it. We have developed a  simple model of a delayed scatterer by modifying the random function that describes a homogeneous extended scatterer. 
Our model allows us to 
obtain a relation between the deterministic parameters of the target model and statistical moments of the SAR image. We assume a regular shape of the antenna trajectory, and our model targets are localized in at least one space-time coordinate; this permits analytical formulation for statistical moments of the image. 
The problem of reconstruction of coordinate-delay reflectivity function is reduced to that of discrimination between instantaneous and delayed scatterers; for the latter problem, the maximum likelihood based image processing procedure has been developed.  
We perform Monte-Carlo simulation and evaluate performance of the classification procedure for a simple dependence of scatterer reflectivity on the delay time.
\end{abstract}


\setcounter{footnote}{0}

\section{Introduction}

Detection of targets with delayed response, or the so-called dispersive targets, can provide valuable information for the interpretation of the observed scene in synthetic aperture radar (SAR) imaging. Man-made objects often exhibit delayed response, and the characteristics of reflectivity in the ``delay'' coordinate depend on the scale, internal structure, and material of the target \cite{chen-2002b,medina-02,albanese-13,cheney-13,sotirelis-13,ferrara-2017}.

Two major obstacles to retrieving the delay information from radar signals are the range-delay ambiguity and speckle. Radar images are built by processing the signals that have been emitted by the radar antenna, scattered by a target, and then received  by either the same or a different antenna. The received signal is a function of time, which is a single scalar variable. The central assumption of the signal processing algorithms (i.e., the algorithms that convert the signal into an image, which is a function of two target coordinates) is  that the travel time of a signal is proportional to the travel distance, given that the propagation speed is constant. This relation no longer holds if reflection at the target involves some delay. 
When this delay exceeds the travel time between the adjacent image pixels, the delayed return contributes to the instantaneous returns from the pixels at larger distances from the antenna compared to the pixel  containing the delayed target. Hence, objects with delayed return will appear in the images as streaks in the range direction.  
When there is only one radar signal involved, there is no possibility to distinguish, without additional information, between the delayed response from an object and an immediate return from another object at a larger distance. 

The range-delay ambiguity  can, in principle, be resolved if we consider the reflection of multiple signals that impinge on the target  from different directions, as done in SAR imaging. Indeed, the distance  between the antenna and various parts of the target is  a function of the observation angle. At the same time, the delayed response is typically determined by the internal composition and/or geometry of the target and does not depend on the observation angle. 
This difference in the properties of the two types of reflected signals is exploited  in \cite{ferrara-2017} for building a model for SAR imaging of the  targets with the reflectivity function that varies in space and may also involve a delayed component.

The resulting procedure, however, appears to have insufficient  sensitivity and low resolution in the delay variable. For narrow and moderate apertures, which are the most popular SAR acquisition modes, the range-delay ambiguity is not very well resolved. Additionally, the speckle effect \cite{goodman-76,goodman-84}, which is common in SAR images, complicates the detection of small variations of image intensity \cite{oliver-98}. This effect is often described in statistical framework where the image pixels are random variables with certain probability distributions. 
Hence, individual pixel values in the image are not a reliable source of information about the target reflectivity as a function of the coordinates and delay. 
In other words, if the difference between the images of two deterministic targets, instantaneous and delayed,  is small in the first place, then, on a random background due to speckle, these targets may become indistinguishable.

In this work, we address the challenges due to the presence of speckle and range-delay ambiguity by means of accumulating and processing the redundant data. In particular, we adopt a statistical approach for the description of the scattering process and imaging and  use multiple delay measurements 
in order to mitigate the stochastic effects. 
We propose an ``incoherent'' scatterer model, which is an extension of 
the commonly used concept of ``uniform delta-correlated background,'' to describe different scatterer types, including those with delayed response. 
Whereas the standard objective for SAR is to reconstruct the scattering characteristics of the target, 
our aim is rather to detect a delayed return immersed into an instantaneous background. This can be thought of as ``lowering the bar'' for the output of the inverse problem as compared to~\cite{ferrara-2017}. The 
gain here 
is that our approach helps increase the robustness of discrimination between instantaneous and delayed targets. We will see,  however,  that an unfavorable combination of the target and system parameters can still make a reliable discrimination impossible. 
Accordingly, we will assess the performance of our procedure by the percentage of incorrect discriminations between different scene types.

The literature on conventional SAR is substantial, see, e.g., \cite{cumming-05,cheney-09}, as well as \cite[Chapter~2]{sarbook}.
An extensive  review on SAR imaging of non-instantaneous targets can be found in the recent article~\cite{ferrara-2017}. The specific approach proposed in \cite{ferrara-2017} for handling the delayed returns is based on the coordinate-delay imaging operator (see Section~\ref{sec:ssttW}). An alternative to that is sub-banding, see, e.g., \cite{albanese-13}. The latter technique  involves splitting the available bandwidth into several sub-bands and building individual SAR images in each sub-band. This, of course,   decreases the  resolution of the entire image. Besides, long and gradual response delays, such as the one due to a cavity in~\cite{ferrara-2017}, have a very narrow manifestation in the frequency domain. Hence, sub-banding may remain inefficient until the resolution becomes as low as the streak length. 
Another class of approaches, see \cite{borden-98} and~\cite[Chapter~4]{chen-2002b}, have the delayed response  parametrized by means of a sum of several fixed-frequency waveforms followed by either spectral filtering or an optimization procedure.  This method should work well if the delayed return is highly coherent, i.e.,  concentrated in a few waveforms, each having a bandwidth much smaller than that of the signal. The effect of speckle is not accounted for in either \cite{borden-98} or~\cite[Chapter~4]{chen-2002b} though. 

The current paper starts with the analysis of the coordinate-delay imaging operator in Section~\ref{sec:ssttW}. Statistical models of the scene components, such as the background, the delayed target, and the instantaneous inhomogeneous target, are introduced in Section~\ref{sec:delayedreflectivity}. 
The discrimination procedure is described in Section~\ref{sec:discrimination}, while the Monte-Carlo simulation to assess the quality of discrimination is described in Section~\ref{sec:MonteCarlo}. The results and discussion are presented in Sections~\ref{sec:ssttResults} and~\ref{sec:ssttDiscussion}, respectively. Some technical details are provided in \ref{sec:FGH}. 

\section{Coordinate-delay imaging operator}
\label{sec:ssttW}

\subsection{Instantaneous and delayed scattering and range-delay ambiguity}
\label{sec:rangedelay}

In  monostatic SAR imaging, multiple signals are emitted by the radar antenna, scattered about the target, and received by the same antenna. The propagation of a 
scalar field $u^\incid(t,\z)$ due to forcing $P(t)$ at 
the point~$\x$ is described by 
$$
	u^\incid(t,\z) = \frac{1}{4\pi R_\z} P\Big(t-\frac{R_\z}{c}\Big), 
$$
where $R_\z=|\z-\x|$. %
Following~\cite{ferrara-2017} and~\cite[Chapter~6]{cheney-09}, we describe the relation between the incident field $u^\incid$ and scattered field $u^\sct$  as convolution in time: 
\begin{equation}
\label{eq:linscattering}
\begin{aligned}
	u^\sct(t,\z) 
&=\> 
	\int_0^\infty u^\incid(t-t_\z,\z)\nu(t_\z,\z)\,dt_\z
\\ &=\>
	\int_0^\infty \frac{1}{4\pi R_\z} P\Big(t-\frac{R_\z}{c} - t_\z\Big) \nu(t_\z,\z)\,dt_\z
	.
\end{aligned}
\end{equation}
In this formulation, the reflectivity function $\nu(t_\z,\z)$ describes the delayed response of a linear material to the incident field. The lower limit of integration in (\ref{eq:linscattering}) is set to zero due to the causality, and the interaction of the scattered field with the target is neglected following the first Born approximation. Hereafter, we will assume that the integrals of the type that appear in (\ref{eq:linscattering}) are finite.

For standard SAR  where all the targets are  instantaneous, formula (\ref{eq:linscattering}) simplifies and the relation between the scattered and emitted field becomes:
\begin{equation}
\label{eq:instantaneous}
	u^\sct(t,\z)=\nuinst (\z) u^\incid(t,\z)
\Longrightarrow 
	\nu(t_\z,\z)=\nuinst(\z)\delta(t_\z), 
\end{equation}
where $\delta$ is the Dirac delta function. 

In this work, we take the pulse receive location to be the same as its transmit location, i.e., $\x$. 
Propagating the scattered signal~\eqref{eq:linscattering} back to~$\x$ from all scattering locations~$\z$, we obtain  
\begin{equation}
\label{eq:lippmann}
	u^\sct_\x(t) \equiv u^\sct(t,\x) = \int_0^\infty dt_\z \int d\z\, \frac{1}{16\pi^2 R_\z^2} \nu(t_\z,\z) P\Big(t-\frac{2R_\z}{c}-t_\z\Big) 
	.
\end{equation}
In what follows, we will assume that the target is observed from the distance much larger than its size. This allows us to disregard the dependence of~$R_\z^{-2}$ on~$\z$ and, subsequently, incorporate $\frac{1}{16\pi^2 R_\z^2}$ into~$\nu$ as a constant factor. 
Moreover, the motion of the antenna during the pulse transmission and reception causes the Doppler shift of the signal frequency; we assume that this effect can be disregarded as well (this is the so-called start-stop approximation; its validity has been explored in \cite[Chapter~6]{sarbook}). 

The range-delay ambiguity is easiest to understand in the 1D case where $\x = x$, $\z \equiv z$, and $R_z = R+z$,  $R=\text{const}$. 
Then, it appears fundamentally  impossible to unambiguously reconstruct a function of two arguments $\nu(t_z,z)$ from the function of a single argument~$u^\sct_x(t)$. Indeed, the substitution
\begin{equation}
\label{eq:rdambiguity_z}
	\nu(t_z,z) \mapsfrom \nu'(t_z,z) = \nu(t_z,z) + f\Big(\frac{2R_z}{c} + t_z\Big)g(t_z),
\end{equation}
where
$
	\int_0^\infty g(t_z)\,dt_\z=0
$ 
and~$f(t_z)$ is an integrable function, does not affect $u^\sct_x(t)$ given by~\eqref{eq:lippmann}, as  one can see by changing the integration variables: $(t_z,z) \mapsto (t_z,t_z+2z/c)$. Hence, the inversion  $u^\sct_x(t) \mapsto \nu(t_z,z)$ cannot be unique. 

A single-pulse coordinate-delay image is formed by the application of a matched filter to the received signal (see, e.g. \cite{ferrara-2017}):
\begin{equation}
\label{eq:Isignal}
\begin{aligned}
	I_\x(t_\y,\y) 
&=\> 
	\int \underbrace{\overline{P\Big(t-\frac{2R_\y}{c}-t_\y\Big)}}_\text{matched filter} 
	u^\sct_\x(t) \,dt,
\end{aligned}
\end{equation}
where $R_\y = |\y-\x|$
and the overbar means complex conjugation. 
Substituting~\eqref{eq:lippmann} into (\ref{eq:Isignal}) and changing the order of integration, we obtain a convolution expression for the image:
\begin{equation}
\label{eq:Ixddd}
\begin{aligned}
	I_\x(t_\y,\y) 
&=\> 
	\int_0^\infty dt_\z \int \, d\z\, \nu(t_\z,\z) 
	\underbrace{\int dt \,  \overline{P\Big(t-\frac{2R_\y}{c}-t_\y\Big)}
	P\Big(t-\frac{2R_\z}{c}-t_\z\Big)}_{W_\x(t_\y,\y;t_\z,\z)} 
	.
\end{aligned}
\end{equation}
In~\eqref{eq:Ixddd}, 
the kernel of the transformation $ \nu(t_\z,\z) \mapsto I_\x(t_\y,\y)$, or the imaging kernel, is the point spread function (PSF) $W_\x(t_\y,\y;t_\z,\z)$.  
Changing the integration variable in the innermost integral of \eqref{eq:Ixddd}, we can show that $W_\x$~is  a function of one argument:
\begin{equation}
\label{eq:Wxambig}
	W_\x(t_\y,\y;t_\z,\z) \equiv W_\x\Big(t_\y-t_\z+\frac{2R_\y-2R_\z}{c}\Big).
\end{equation}
Hence, $I_\x(t_\y,\y)$ 
turns out to be 
a function of only one argument, $(t_\y + 2R_\y/c)$, as well. For example, 
\begin{equation}
\label{eq:Ixambig}
	I_\x(t_\y,\y) = I_\x(0, \y')
\quad\text{whenever}\quad
	|\y'-\x| = |\y-\x| + \frac{ct_\y}{2}
	.
\end{equation}
While \eqref{eq:rdambiguity_z} can be seen as a manifestation of the range-delay ambiguity in the target coordinates, i.e., $t_\z$ and~$\z$, formula~\eqref{eq:Ixambig} describes the same effect in terms of the image coordinates $t_\y$ and~$\y$. 

For standard SAR and instantaneous targets, see~\eqref{eq:instantaneous}, formula~\eqref{eq:Ixddd} becomes
\begin{equation}
\label{eq:stdIx}
	I_\x(\y) = \int d\z\,  \nuinst(\z) \underbrace{ \int dt \,  \overline{P\Big(t-\frac{2R_\y}{c}\Big)}
	P\Big(t-\frac{2R_\z}{c}\Big)}_{W_\x(\y,\z) \equiv W_\x(R_\y-R_\z)},
\end{equation}
and the PSF no longer has the time arguments. 

\subsection{Coordinate-delay SAR image and the kernel of the imaging operator}
\label{sec:ssttKernel}

The range-delay ambiguity can be resolved by interrogating the target from different antenna positions. In other words, the observation direction must span a certain sufficiently wide interval. 
Wide-angle SAR imaging is described, e.g., in  \cite{moses-04}.  
The ultimate case of the so-called (full) circular SAR, where the observation platform makes a circle around the target during the image acquisition, is presented in \cite{ishimaru-99-ACF,moore-07}. However, for wide-angle SAR   one can no longer  assume that the reflectivity does not depend on the direction, as in \eqref{eq:linscattering}. In this work, we rather want formula \eqref{eq:linscattering} to hold so that $\nu$ does not depend on~$\x$, a condition sometimes called angular coherence, see, e.g., \cite{moses-04,kim-03,bleszynski-13,varshney-06}. This implies that there is a dominant observation direction. On the other hand, for detecting a delayed response the span of observation angles may not be too narrow, as we will  see in Section~\ref{sec:imagesInhom},  formula \eqref{eq:kappazetamax}.

\begin{figure}[ht]
\centerline{\includegraphics[width=5in,clip=true]{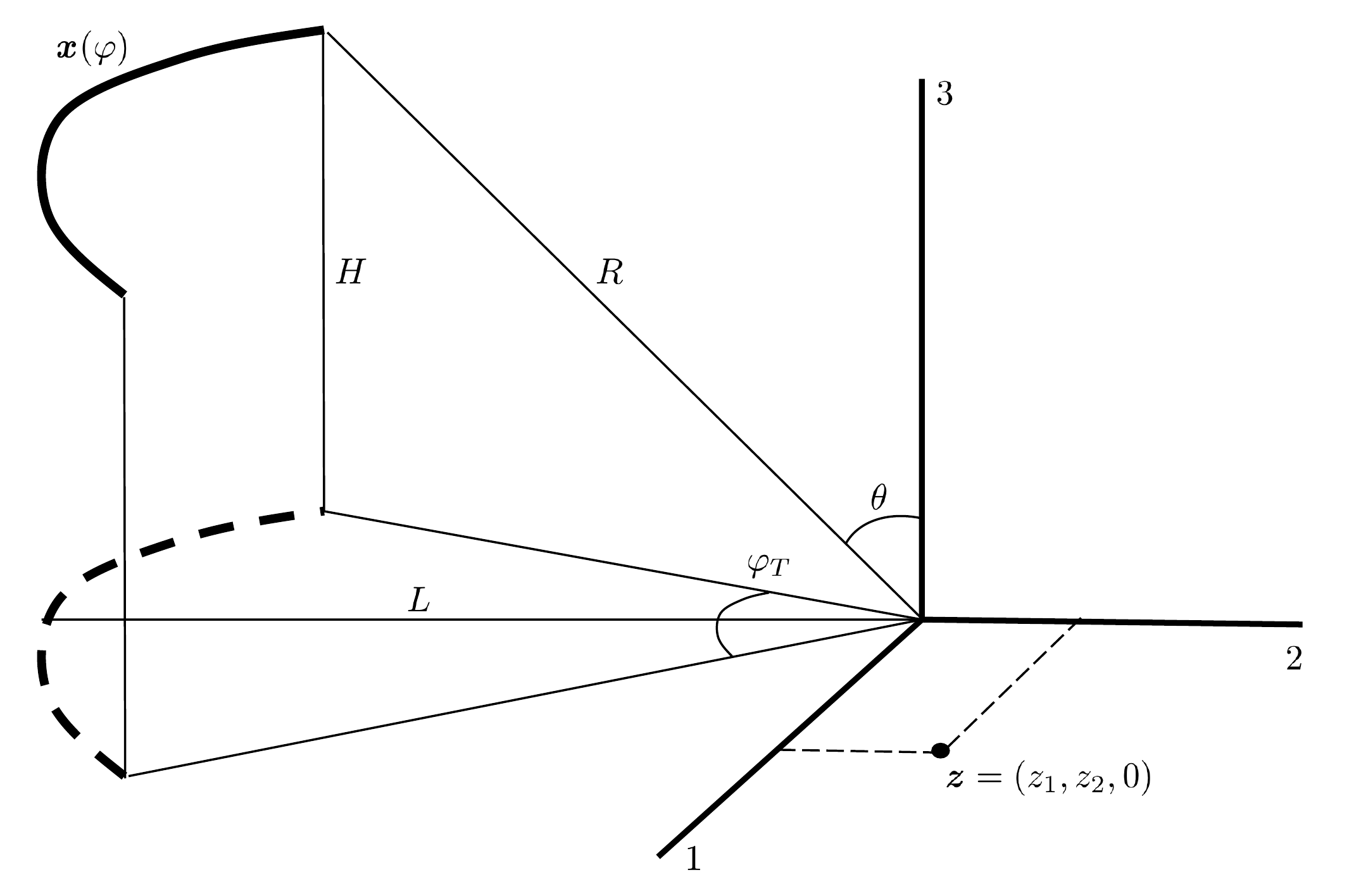}}
\caption{Geometry of the problem.
} 
\label{fig:sstt_geometry}
\end{figure}

We take $x_2$ as the horizontal coordinate aligned with the dominant observation direction, $x_1$ normal to it and also horizontal, and~$x_3$ vertical, as shown in Figure~\ref{fig:sstt_geometry}. The direction $x_2$ will be referred to as range, and $x_1$
as azimuth or cross-range. 
The antenna trajectory at a reference distance~$R$ from the target is specified as an arc of a circle:
\begin{equation}
\label{eq:xvarphi}
	\x = \x(\varphi) =  \veclist{-L\sin\varphi, -L\cos\varphi,H}
,\quad	
	|\varphi|\le \phiT/2
	,
\end{equation}
where $\varphi$ is the aspect angle and $\phiT$ defines the synthetic aperture or, more precisely, its angular width. In addition to that, $L= R\sin\theta$ in formula (\ref{eq:xvarphi}) is the circle radius, $H = R\cos\theta$ is the platform altitude, and $\theta$ is the incidence angle, see Figure~\ref{fig:sstt_geometry}. The coordinates associated with the target and  image will be denoted by $\z=(z_1,z_2,z_3)$ and $\y=(y_1,y_2,0)$, respectively.
Moreover, the scattering will be assumed to occur only on the surface of the target, i.e., on the plane $z_3=0$, which is a common assumption in SAR.
Hence, throughout this paper we will consider
\begin{equation}
\label{eq:z3D2D}
	\z=(z_1,z_2,0)
\quad\text{and}\quad
	\y=(y_1,y_2,0)
	.
\end{equation}
Accordingly, the reflectivities 
$\nu$ and~$\nuinst$,
see \eqref{eq:linscattering} and~\eqref{eq:instantaneous}, will depend only on two spatial coordinates
(see \cite{a71e} and~\cite[Chapter~7]{sarbook} for additional detail).

We will not consider either very steep or very gradual incidence, which implies  $\sin \theta = {\cal O}(1)$ and $\cos\theta = {\cal O}(1)$. The presence of a dominant observation direction  requires $\phiT \ll 1$. 
For~$R^\varphi_\z \bydef |\z-\x(\varphi)|$  we have:
$$
	R^\varphi_\z = \big(R^2 + z_1^2+z_2^2+ 2L(z_1\sin\varphi+z_2\cos\varphi)\big)^{1/2}
	.
$$ 
Assuming that the entire target area of interest is near the origin of the coordinate system, $|\y|,|\z| \ll R$ (see Figure~\ref{fig:sstt_geometry}), we simplify the previous expression as follows: 
\begin{equation}
\label{eq:Rphiz}
	R^\varphi_\z \approx R + \sin\theta(z_1\sin\varphi+z_2\cos\varphi)
	.
\end{equation}
The expression for~$R^\varphi_\y \bydef |\y-\x(\varphi)| $ is obtained similarly.

Define the total image as the sum of single-pulse images \eqref{eq:Isignal}:
\begin{equation}
\label{eq:Isum}
	I(t_\y,\y) 
	= \sum_{n=-N/2}^{N/2} 
	\int \overline{P\Big(t-\frac{2R^n_\y}{c}-t_\y\Big)} 
	u^\sct_{\x^n}(t) \,dt
	= \sum_{n=-N/2}^{N/2}
	I_{\x^n}(t_\y,\y) 
	,
\end{equation}
where~$N$ is the total number of the pulse transmit-receive locations and 
$\x^n$ are given by~\eqref{eq:xvarphi} for $\varphi = \varphi_n = n\phiT/N$. 
When $\{\varphi_n\}$ are sufficiently dense (see \cite[Section~2.4.2]{sarbook} or~\cite[Section~4.5.3]{cumming-05} for detail), the sum in~\eqref{eq:Isum} can be replaced with an integral over~$\varphi$. Then, using~\eqref{eq:Ixddd}, we obtain: 
\begin{equation}
\label{eq:Iconvconv}
	I(t_\y,\y) 
	\approx \frac{N}{\phiT} \int_{-\phiT/2}^{\phiT/2} I_{\x(\varphi)} (t_\y,\y) \,d\varphi
	= \int_0^\infty dt_\z \int \, d\z\, \nu(t_\z,\z) 
	W(t_\y,\y;t_\z,\z),
\end{equation}
 where 
\begin{equation}
\label{eq:Wttyz}
	W(t_\y,\y;t_\z,\z) = \frac{N}{\phiT} \int_{-\phiT/2}^{\phiT/2} d\varphi \int dt \,
	 \overline{P\Big(t-\frac{2R^\varphi_\y}{c}-t_\y\Big)}
	P\Big(t-\frac{2R^\varphi_\z}{c}-t_\z\Big)
	.
\end{equation}

The most common   SAR signal is a chirp:
\begin{equation}
\label{eq:chirp}
	P(t)=A(t)\exp({-i\omega_0 t}),\quad\text{where}\quad A(t)=\chi_\tau(t)\exp({-i\alpha t^2})
\end{equation}
and $\chi_\tau$ is the indicator function:
\begin{equation}
\label{eq:indicator}
	\chi_\tau(t)=\begin{cases} 1, & t\in[-\tau/2,\tau/2],\\
	0, & \text{otherwise}.
	\end{cases}
\end{equation} 
Hereafter, we assume $\alpha > 0$ (``upchirp'') so that the bandwidth $B=2\alpha \tau$ is positive; the case $\alpha < 0$ can be treated similarly. We also assume a narrowband signal with a high time-bandwidth product: $B \ll \omega_0$ and $B\tau \gg 1$, which is also common for SAR.   
Using \eqref{eq:Rphiz} and~\eqref{eq:chirp}--\eqref{eq:indicator}, we transform~\eqref{eq:Wttyz} into  
\begin{equation}
\label{eq:Wintint}
	W(t_\y,\y;t_\z,\z) = \frac{N}{\phiT} \int_{-\phiT/2}^{\phiT/2} d\varphi \, \exp({-2i\omega_0T^\varphi})	
	\int_{-\tau/2}^{\tau/2} d\tilde t  \,
	\exp({-4i\alpha T^\varphi\tilde{t}})
	,
\end{equation}
where
\begin{equation}
\label{eq:Tphi}
\begin{aligned}
	T^\varphi 
&=\>
	\frac{R_\y^\varphi-R_\z^\varphi}{c} + \frac{t_\y-t_z}{2} 
\\ &=\> 
	\Big(\frac{y_2-z_2}{c}\cos\varphi - \frac{y_1-z_1}{c}\sin\varphi\Big)\sin\theta + \frac{t_\y-t_\z}{2} 
	,
\\
	\tilde t 
&=\> 
	t - \frac{R_\y^\varphi - R_\z^\varphi}{c} - \frac{t_\y-t_\z}{2}
	.
\end{aligned}
\end{equation}
In~\eqref{eq:Wintint}, we  made a common simplification by disregarding the dependence of the integration limits on~$\y$ and~$\z$ for signals with~$B\tau \gg 1$, see \cite[Chapter~2]{sarbook}.

Note that, $T^\varphi$ is a function of $t_\y$, $\y$, $t_\z$, $\z$, and~$\varphi$, while~$W$ is  defined via an integral that involves $T^\varphi$. In particular, the interior integral in~\eqref{eq:Wintint} can be evaluated as follows: 
\begin{equation*}
\begin{aligned}
	W_\rng 	
&\equiv\> 
	W_\rng(t_\y,\y;t_\z,\z; \varphi)
\\ &=\> 
	\int_{-\tau/2}^{\tau/2} d\tilde t  \,
	\exp({-4i\alpha T^\varphi\tilde{t}})
= %
	\tau \sinc(B T^\varphi)
	,
\end{aligned}
\end{equation*}
where $\sinc\xi \bydef \sin\xi/\xi$. Thus, 
\begin{equation}
\label{eq:WintWr}
	W(t_\y,\y;t_\z,\z) = \frac{N}{\phiT} \int_{-\phiT/2}^{\phiT/2}  \exp({-2i\omega_0T^\varphi}) W_\rng(t_\y,\y;t_\z,\z; \varphi) \,d\varphi
	.
\end{equation}

For $T^\varphi$ given by~\eqref{eq:Tphi}, we will take the Taylor expansion of the trigonometric functions of~$\varphi$ about zero and explore the effect of its first three terms on expression~\eqref{eq:WintWr}. 
If we retain the zeroth-order term only, i.e.,  $\cos \varphi \approx 1, \sin \varphi \approx 0$, then the imaging kernel \eqref{eq:Wintint} does not depend on the cross-range coordinates $y_1$ and~$z_1$ at all. Then, the radar will not be able to reconstruct any variation of $\nu$ in the cross-range direction, i.e., will provide no azimuthal resolution.

Expanding~$T^\varphi$ up to the linear term, i.e., $\cos\varphi \approx 1$, $\sin \varphi \approx \varphi$, we obtain:
\begin{align}
\nonumber
	\Wlin(t_\y,\y;t_\z,\z) 
&=\> 
	\exp({-2i\omega_0T^0}) \frac{N}{\phiT} \int_{-\phiT/2}^{\phiT/2}  \exp\big({2i\kOtheta(y_1-z_1)\varphi}\big) 	  W_\rng \,d\varphi
\\ \nonumber 
&=\> 
	\tau \exp({-2i\omega_0T^0}) \frac{N}{\phiT} \int_{-\phiT/2}^{\phiT/2}  \exp\big({2i\kOtheta(y_1-z_1)\varphi}\big)  
\\ \label{eq:Wlinphi}
& \phantom{
      =\> 
	\tau \exp({-2i\omega_0T^0}) \frac{N}{\phiT} \int_{-\phiT/2}^{\phiT/2}  
}		
	\cdot \sinc (B \Tlinphi) \,d\varphi, 
\end{align}
where 
\begin{equation}
\label{eq:T0}
\begin{split}
	T^0 = T^\varphi \Big|_{\varphi=0} = 	\frac{y_2-z_2}{c} \sin\theta + \frac{t_\y-t_\z}{2} 
	,
\\ 
	\Tlinphi  = T^0 -  \sin\theta \frac{y_1-z_1}{c} \varphi
,\quad 
	\kOtheta = \frac{\omega_0}{c}\sin\theta
	.
\end{split}
\end{equation}
Unlike previously, the imaging kernel $\Wlin$ of \eqref{eq:Wlinphi} does depend 
 on the cross-range coordinates. Moreover, the range coordinates~$y_2$ and~$z_2$ appear only in combination with~$(t_\y-t_\z)$ as in~$T^0$, see~\eqref{eq:T0}. This, in particular, means:
\begin{equation}
\label{eq:Wambig}
\begin{aligned}
	\Wlin(t_\y,\y;t_\z,\z) 
&=\> 
	\Wlin\Big(0, \y + \frac{ct_\y}{2\sin\theta}\uvector_2; \; t_\z,\z\Big) 
\\ &=\> 	
	\Wlin\Big(t_\y,\y; \; 0, \z + \frac{ct_\z}{2\sin\theta}\uvector_2\Big)
	,
\end{aligned}
\end{equation}
and, due to~\eqref{eq:Iconvconv}, 
\begin{equation}
\label{eq:Iambig}
	\Ilin(t_\y,\y) = \Ilin\Big(0, \y + \frac{ct_\y}{2\sin\theta}\uvector_2\Big)
	,
\end{equation}
where~$\uvector_2$ is a unit vector in the range direction (cf.\ formula \eqref{eq:Ixambig}). We see that the range-delay ambiguity is not resolved if we retain only the linear term with respect to $\varphi$ in the expansion of~$T^\varphi$ (in this regard, expressions \eqref{eq:Wambig} and \eqref{eq:Iambig} are similar to their single-pulse counterparts~\eqref{eq:Wxambig} and~\eqref{eq:Ixambig}, respectively).

Replacing~$\Tlinphi$ with~$T^0$ under the $\sinc$ in~\eqref{eq:Wlinphi} (but not in the exponent in~\eqref{eq:WintWr}, where the factor in front of~$\varphi$ is $2\omega_0/B \gg 1$~times bigger) allows us to integrate over~$\varphi$ and obtain the factorized expression:%
\footnote{
It has been shown in~\cite[Chapter~2]{sarbook} and~\cite{a70e} that if we retain the dependence of~$W_\rng$ on~$\varphi$ (i.e., if we don't replace~$T^\varphi$ with~$T^0$ under the $\sinc$ in~\eqref{eq:Wlinphi}, which is equivalent to replacing~$W_\rng$ with~$W_\rng^0$ in~\eqref{eq:WintWr}), then the difference between the expressions~\eqref{eq:WintWr} and~\eqref{eq:W3factors} is small, on the order of $N\tau B/\omega_0$ by absolute value, or about $B/\omega_0 $ in relative terms (this difference is called the factorization error). 
}
\begin{equation}
\label{eq:W3factors}
	\Wlin(t_\y,\y;t_\z,\z) \approx  \exp({-2i\omega_0T^0}) \cdot 
	\underbrace{ 
		N \sinc \big(\kOtheta \phiT (y_1-z_1)\big)
	}_{W_\az}
	 \cdot 
	\underbrace{ 
		\vphantom{N \sinc \big(\kOtheta \phiT (y_1-z_1)\big)}
		\tau \sinc(BT^0)
	}_{W_\rng^0}
	,
\end{equation}
where $W_\az$ and~$W_\rng^0$ are the azimuthal and range factors of the imaging kernel, respectively. 
The case of a standard SAR (see also~\eqref{eq:stdIx}) corresponds to 
$\nu$~given by~\eqref{eq:instantaneous} and $t_\y=0$, which turns formulae~\eqref{eq:Iconvconv} and~\eqref{eq:W3factors} into the following:
\begin{equation}
\label{eq:Istd}
	I_\text{std}(\y) 
	\bydef \Ilin(0,\y)
	= \int d\z\,  \nuinst(\z) \Wlin(0,\y;0,\x)
	\bydef \int d\z\,  \nuinst(\z) W_\text{std}(\y,\z)
	,
\end{equation}
where 
\begin{equation}
\label{eq:Wstd}
\begin{aligned}
	W_\text{std}(\y,\z) 
=&\> 
	N\tau  \exp\big({-2i\omega_0 \sin\theta(y_2-z_2)/c}\big)
\\ &\> \cdot 
	\sinc \big(\kOtheta \phiT (y_1-z_1)\big) 
	\sinc\Big(B\frac{y_2-z_2}{c}\sin\theta\Big)
\end{aligned} 
\end{equation}
is the corresponding point spread function, 
and 
the three factors on the right hand side of~\eqref{eq:Wstd} (excluding the factor $N\tau$) define the fast phase, cross-range response, and range response, respectively. 
The function~$W_\text{std}(\y,\zd)$ is proportional to the image of an instantaneous point scatterer:
\begin{equation}
\label{eq:spacedelta}
	\nuinst(\z) = A \delta(\z-\zd)
\quad\text{where}\quad
	\zd = (\zdO,\zdT,0),
\end{equation}
and hence the resolution in azimuth $\Delta_\az$ and resolution in range~$\Delta_\rng$ can be defined as  semi-width of the main lobe of the corresponding sinc term. This yields: 
$$
|W_\text{std}(\y,\z)| \sim 
	\Big| \sinc \Big(\pi\frac{y_1-z_1}{\Delta_\az} \Big) \sinc\Big(\pi\frac{y_2-z_2}{\Delta_\rng}\Big)
	\Big|
	,
$$
where 
\begin{equation}
\label{eq:DeltaRA}
	\Delta_\az = \pi \frac{1}{\kOtheta\phiT} 
\quad\text{and}\quad 
	\Delta_\rng = \pi \frac{c}{B\sin\theta}
	.
\end{equation}

The simplified expression~\eqref{eq:W3factors} allows us to attribute the range-delay ambiguity to~$W_\rng^0$, which becomes the range factor of the imaging kernel in the case of a standard SAR, see \eqref{eq:Wstd}. 
As the linearized approximation of~$T^\varphi$ given by \eqref{eq:T0} is found insufficient to resolve the range-delay  ambiguity, we bring along the  quadratic terms in the expansion of~$\cos\varphi$ and $\sin\varphi$, which yields:
\begin{equation}
\label{eq:cosTaylor2}
	\cos\varphi \approx 1-\frac{\varphi^2}{2}
,\quad
	\sin\varphi \approx\varphi
	.
\end{equation}
Using the same factorized approximation as in~\eqref{eq:W3factors}, we obtain:
\begin{equation}
\label{eq:Wintsq}
\begin{aligned}
	W(t_\y,\y;t_\z,\z) 
\approx&\> 
	\exp({-2i\omega_0T^0}) W_\rng^0 \frac{N}{\phiT} 
\\ &\> \cdot 	
	\int_{-\phiT/2}^{\phiT/2}  \exp\big({2i\kOtheta(y_1-z_1)\varphi}\big)
	\exp\big({i\kOtheta(y_2-z_2)\varphi^2}\big)
	 \,d\varphi
	 .
\end{aligned} 
\end{equation}

\begin{figure}[ht]
\hspace{-1.3cm}
	\includegraphics[width=6.5in]{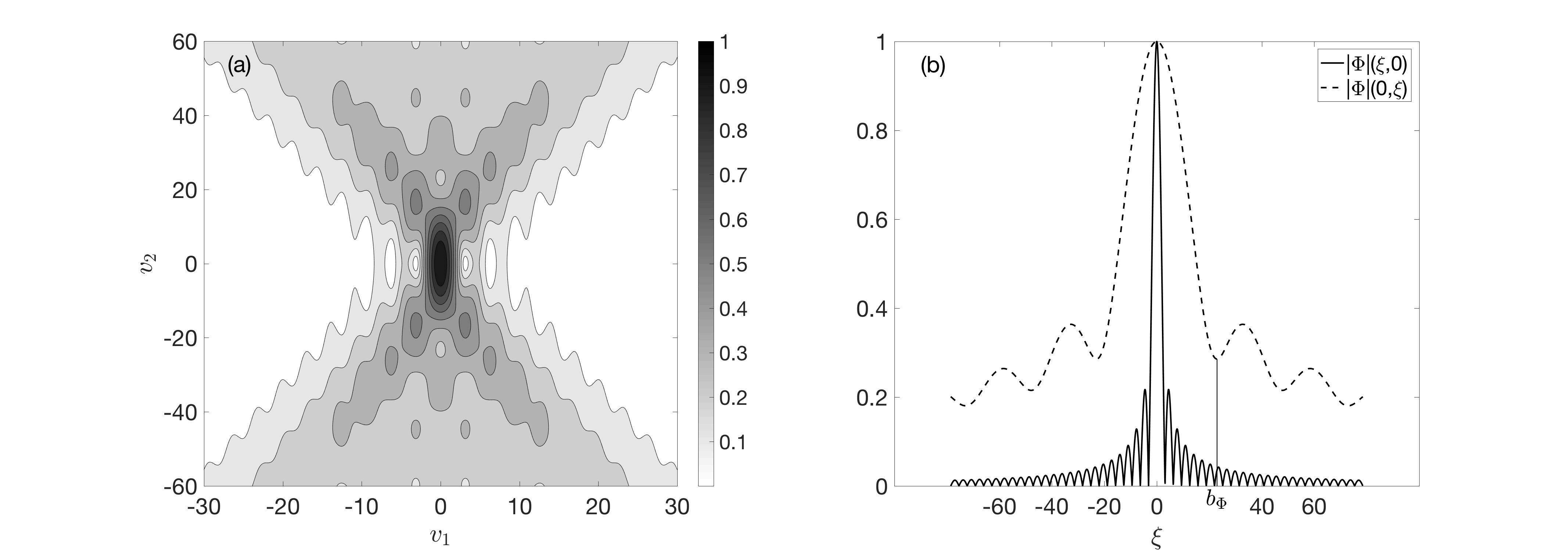} %
\vspace{-0.4cm}
\caption{Left: Contour plot of $|\Phi|(v_1,v_2)$; Right: marginal functions~\eqref{eq:PhiPhi} of $|\Phi|(v_1, v_2)$. The half-width of the main lobe of~$|\Phi|(\xi,0)$ is~$\pi$, whereas that of~$|\Phi|(0,\xi)$ is $\bPhi \approx 23$.  
} 
\label{fig:Phimarginals}
\end{figure}

Similarly to \cite[formula~(59)]{garnier-16}, 
we introduce a function of two variables:
\begin{equation}
\label{eq:Phidef}
	\Phi(v_1,v_2) \bydef \int_{-1/2}^{1/2} \exp({2iv_1s})\exp({iv_2s^2})\,ds
	.
\end{equation}
It is easy to show that 
\begin{equation}
\label{eq:PhiPhi}
	\Phi(v_1,0) = \sinc v_1
\quad\text{and}\quad
	\Phi(0,v_2) = \frac{C(t) + i\sign(v_2)S(t)}{t},
\end{equation} 
where $t = |v_2|^{1/2}(2\pi)^{-1/2}$ and 
$C(t)$ and $S(t)$ are the Fresnel integrals \cite{NIST:DLMF}. The absolute value of~$\Phi$, as well as the marginal functions~\eqref{eq:PhiPhi}, are plotted in Figure~\ref{fig:Phimarginals}. Both marginal functions have their peaks when the corresponding argument is zero. In doing so,  the main lobe of~$\Phi(v_1,0)$ is clearly delineated by the zeros at~$|v_1|=\pi$, yet the minima of~$|\Phi(0,v_2)|$ at~$|v_2| = \bPhi \approx 23$ %
appear quite ``shallow.'' 
For $|v_2| \gtrsim 1$, the stationary phase analysis  yields 
\begin{equation}
\label{eq:condv1v2}
	|v_1| \le |v_2|/2 
\end{equation}
as the condition  for the stationary point of the integral in~\eqref{eq:Phidef} to be within the integration limits. When condition (\ref{eq:condv1v2}) is not satisfied, the value of the integral is small, as indicated by the white areas in the left panel of Figure~\ref{fig:Phimarginals}. 

With the help of~\eqref{eq:Phidef}, formula~\eqref{eq:Wintsq} can be expressed as 
\begin{equation}
\label{eq:WviaPhi}
	W(t_\y,\y;t_\z,\z) 
	= 
	\exp({-2i\omega_0T^0}) \cdot N  \Phiarg{(y_1-z_1)}{(y_2-z_2)} \cdot W_\rng^0
	,
\end{equation}
where $W_\rng^0 = \tau \sinc (BT^0) $ as in~\eqref{eq:W3factors}, with $T^0$ defined by~\eqref{eq:T0}. This form of the imaging kernel will be used throughout the rest of this paper.

\subsection{Properties of the coordinate-delay imaging operator} 
\label{sec:Wproperties}

We are interested in how well the range-delay ambiguity described in Section~\ref{sec:rangedelay} is resolved when imaging with the help of the kernel~\eqref{eq:WviaPhi}. Formally, $W$ is a function of six scalar arguments, although the particular form~\eqref{eq:WviaPhi} indicates that we can reduce its set of arguments to only three independent variables: $(y_1-z_1)$, $(y_2-z_2)$, and~$(t_\y-t_\z)$. This should be  expected because, given some constant $\td$ and $\zd$, the function 
$W(t_\y,\y;\td,\zd)$ is proportional to the image due to a ``space-time point scatterer:'' 
\begin{equation}
\label{eq:spacetimedelta}
	\nu(t_\z,\z) = A \delta(t_\z-\td) \delta(\z-\zd), 
\end{equation} 
see~\eqref{eq:Iconvconv}, \eqref{eq:Istd}, and~\eqref{eq:spacedelta}. The maximum of this image amplitude, or, equivalently, the maximum of $|W|$, is achieved at the space-time ``location'' of the scatterer~\eqref{eq:spacetimedelta}, i.e.,  when $(y_1-z_1) = 0$, $(y_2-z_2)=0$, and~$(t_\y-t_\z)=0$. The range-delay ambiguity will be caused by a slow decay of~$|W|$ along certain directions in the space of its arguments. 

Similarly to~\eqref{eq:Wxambig}, the range coordinates are tied with the delay in~$W_\rng^0$: 
\begin{equation}
\label{eq:WRO}
\begin{aligned}
	W_\rng^0 = \tau \sinc (BT^0) 
&=	
	 \tau \sinc\Big[B\Big(\frac{y_2-z_2}{c} \sin\theta + \frac{t_\y-t_\z}{2} \Big)\Big]
\\ &= 
	\tau \sinc\Big[\pi\frac{1}{\Delta_\rng}\Big(y_2-z_2+\frac{t_\y-t_\z}{2\sin\theta}\Big)\Big]
.
\end{aligned}
\end{equation}
However, the range coordinate~$(y_2-z_2)$ stands separate from the delays in the second argument of~$\Phi$, see~\eqref{eq:WviaPhi}. 
Hence, it is 
possible to use the semi-width of the main lobe of the dashed curve in the right panel of Figure~\ref{fig:Phimarginals} to define the ``unambiguous''  resolution size due to the second argument of~$\Phi$: 
\begin{equation}
\label{eq:reslnU}
	\Delta_\una = \frac{\bPhi}{\kOtheta\phiT^2}. 
\end{equation}
It is therefore the interaction of the two factors in (\ref{eq:WviaPhi}), $W_\rng^0$ and~$\Phi$, that will determine how the 
 the range-delay ambiguity manifests itself. Both factors depend on the range coordinates and both decay as $|y_2-z_2| \to \infty$:
\begin{equation}
\label{eq:twoscales}
	|W_\rng^0| \sim \Big|\pi \frac{y_2-z_2}{\Delta_\rng}\Big|^{-1}
, \quad
	|\Phi| \sim  \Big| \frac{\bPhi}{\pi} \frac{y_2-z_2}{\Delta_\una}\Big| ^{-1/2}
	.
\end{equation}
The first estimate \eqref{eq:twoscales} holds assuming that $(t_\y-t_\z)$ is fixed in the argument of $W_\rng^0$, see~\eqref{eq:WRO}. The second estimate \eqref{eq:twoscales} can be obtained by applying the stationary phase formula to~\eqref{eq:Phidef} provided~\eqref{eq:condv1v2} is satisfied.

We introduce the following parameter to describe the ratio between the two scales in~\eqref{eq:twoscales}: 
\begin{equation}
\label{eq:kappadef}
	\kappa \bydef \phiT^2 \frac{\omega_0}{B}
	.
\end{equation}
When $\Delta_\rng \ll \Delta_\una$ (or, equivalently, $\kappa \ll \bPhi/\pi \approx 7$),  the main lobe of~$W$ in the range direction (i.e., as a function of~$(y_2-z_2)$  with $y_1=z_1$ and $t_\y = t_\z$) is determined by~$W_\rng^0$ because the factor~$\Phi$ in~\eqref{eq:WviaPhi} for $B|y_2-z_2|\sin\theta/c \lesssim \pi$ can approximately be replaced with one. In the  opposite case of $\Delta_\rng \gg \Delta_\una$, or $\kappa \gg \bPhi/\pi$, the main lobe of~$W$ in the range direction for $B|y_2-z_2|\sin\theta/c \lesssim \pi$ is determined by the function~$\Phi$ of~\eqref{eq:Phidef}. 
We will call $\kappa \ll \bPhi/\pi$ and $\kappa \gg \bPhi/\pi$ the narrow-aperture and wide-aperture modes, respectively. 

For a narrow-aperture regime, the resolution of the system in range is due to~$W_\rng^0$. In this case,  $\Phi(\ldots )$ in~\eqref{eq:WviaPhi} turns into $\sinc\big(\kOtheta\phiT(y_1-z_1)\big)$ as $\kappa \to 0$, so~\eqref{eq:WviaPhi} reduces to \eqref{eq:W3factors}. However, the factor $W_\rng^0$ is subject to the range-delay ambiguity, while the range resolution due to~$\Phi(\ldots)$ is much larger: 
\begin{equation}
\label{eq:narrowaperture}
	|y_2-z_2| \gtrsim \frac{\bPhi}{\kOtheta\phiT^2} = \kappa\Delta_\rng \frac{\bPhi}{\pi} \gg \Delta_\rng. 
\end{equation}
The geometry of the narrow-aperture images can be understood as follows. 
As the antenna is far away, see~\eqref{eq:Rphiz}, we define the ambiguity directions:  
\begin{subequations} 
\label{eq:ambigdir}
\begin{align}
\label{eq:ambigdir_z}
	\text{``ambiguity direction in $\z$''} \bydef \Big(t_\z + \frac{2z_2\sin\theta}{c} 
&=\>
	\text{const}, \; z_1 = \text{const} \Big),
\\ 
\label{eq:ambigdir_y}
	\text{``ambiguity direction in $\y$''} \bydef \Big(t_\y + \frac{2y_2\sin\theta}{c} 
&=\>
	\text{const}, \; y_1 = \text{const} \Big). 	
\end{align}
\end{subequations}
The right-hand side of~\eqref{eq:ambigdir_z} specifies a direction in the ``coordinates'' $\big(z_1,z_2, \frac{ct_\z}{2\sin\theta}\big)$ as the intersection of a plane from the family $ t_\z + \frac{2z_2\sin\theta}{c} 
= \text{const}$ with a plane from the family $z_1=\text{const}$, while \eqref{eq:ambigdir_y} does the same for $\big(y_1,y_2, \frac{ct_\y}{2\sin\theta}\big)$. 
A straight line given by such an intersection will be called an ambiguity line. 
The analysis in Sections~\ref{sec:rangedelay} and~\ref{sec:ssttKernel} shows that the images due to the kernel~$\Wlin$ of~\eqref{eq:W3factors} are insensitive to the variations of~$\nu(t_\z,\z)$ that preserve the integral of~$\nu$ along the ambiguity direction in~$\z$, see~\eqref{eq:ambigdir_z}, (\ref{eq:Wambig}), and~\eqref{eq:rdambiguity_z}; at the same time, such images are constant along the ambiguity direction in~$\y$, see~\eqref{eq:ambigdir_y} and~\eqref{eq:Iambig}.

The effect of the quadratic term in~\eqref{eq:cosTaylor2} is controlled by the parameter~$\kappa$ of~\eqref{eq:kappadef}. For example, inequality~\eqref{eq:narrowaperture} means that the image of a point scatterer~\eqref{eq:spacetimedelta} will be stretched along the ambiguity line passing through the ``point'' 
$\big(y_1,y_2, \frac{ct_\y}{2\sin\theta}\big) = \big(\zdO,\zdT, \frac{c\td}{2\sin\theta}\big)$. At the same time, 
the characteristics of this image in the plane~$t_\y = 0$ are still defined by~\eqref{eq:DeltaRA}, 
which means that, unlike in sub-banding \cite{albanese-13}, no formal concession in range resolution is made in the attempt to achieve resolution in the delay variable (see also~\cite{ferrara-2017}).

For the wide-aperture case (large $\kappa$), the range resolution is formally due to~$\Phi$ and there is no ambiguity. However, 
high sidelobes and slow decay of~$|\Phi|$ in its second argument, as per the second equation  \eqref{eq:twoscales} and the right panel of Figure~\ref{fig:Phimarginals}, 
make it difficult to achieve the  
range resolution comparable to $\Delta_\una$ of~\eqref{eq:reslnU}, see also \cite{ferrara-2017,moses-04,moore-07}. 
Hence, in practice the resolution in range is still given by $\Delta_\rng$ of (\ref{eq:DeltaRA}). This may negatively impact the imaging in azimuth as well. 
Indeed, using (\ref{eq:WviaPhi}) we can write:
\begin{equation*}
|W| \propto  \Big| \Phi(v_1,v_2) \cdot W_\rng^0\Big(\frac{v_2}{\kappa}\Big) \Big|,
\end{equation*}
where $v_1=\kOtheta\phiT(y_1-z_1)$ and $v_2=\kOtheta\phiT^2(y_2-z_2)$. Then, taking $|y_2-z_2| \sim \Delta_\rng$ 
in \eqref{eq:condv1v2}, we see that for large $\kappa$ the corresponding azimuthal width of~$\Phi(\ldots)$ shown in the left panel of Figure~\ref{fig:Phimarginals} becomes:
\begin{equation}
\label{eq:ggDeltaAz}
	|y_1-z_1| \sim \frac{1}{2}\frac{\Delta_\rng}{\phiT} = \frac{1}{2} \Delta_\az\kappa \gg \Delta_\az
	.
\end{equation}

To highlight the role of the ambiguity directions  introduced in~\eqref{eq:ambigdir}, we change the corrdinates:
\begin{equation}
\label{eq:etazetapsi}
\begin{aligned}
	\eta 
&=\>
	\kOtheta \phiT(y_1-z_1)
	,
\\
	\zeta
&=\>
	\frac{B}{\omega_0}\kOtheta\Big(y_2 - z_2 + \frac{1}{\sin\theta}\frac{c(t_\y-t_\z)}{2}\Big)
	,
\\
	\psi
&=\>
	\frac{B}{\omega_0}\kOtheta\Big(y_2 - z_2 - \frac{1}{\sin\theta}\frac{c(t_\y-t_\z)}{2}\Big)
	.
\end{aligned}
\end{equation}
In~\eqref{eq:etazetapsi}, $\zeta$ and~$\psi$ are the coordinates across and along the ambiguity lines, respectively. Then, expression~\eqref{eq:WviaPhi} takes the following form: 
\begin{equation}
\label{eq:etazetapsiW}
	W(\eta,\zeta,\psi) 
	= 
	N\tau \exp\Big({-2i\frac{\omega_0}{B}\zeta}\Big) \cdot \Phi\Big(\eta,\kappa\frac{\zeta+\psi}{2}\Big) \cdot \sinc \zeta
	.
\end{equation}
The central peak of~$W$ is well defined along $\eta$ and~$\zeta$ coordinates. However, if we fix~$\eta$ and~$\zeta$, then there is only a slow decay, $\sim \psi^{-1/2}$, in the ambiguity directions in~\eqref{eq:ambigdir}, see the second estimate of (\ref{eq:twoscales}).

\begin{figure}[ht]
\begin{center}
\includegraphics[width=5in]{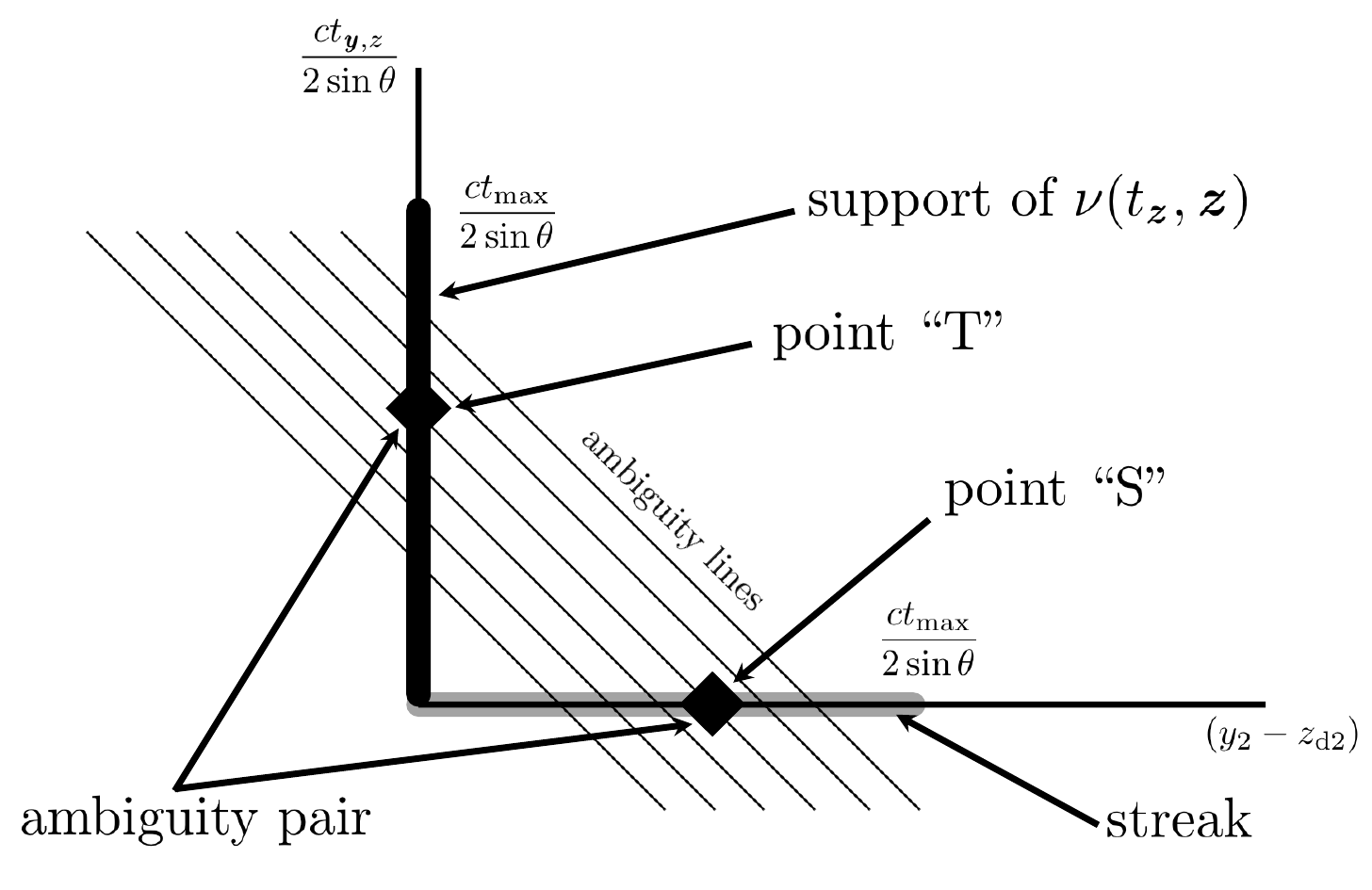}
\vspace{-0.5cm}
\end{center}
\caption{Ambiguity lines \eqref{eq:ambigdir}, ambiguity pair, and the streak due to a delayed scatterer.} 
\label{fig:sstt_streak}
\end{figure}

When an ambiguity line with $\eta = 0$ intersects the planes 
$t_\y=0$ and $y_2=\zdT$ 
(or, in the target coordinates, $z_1=\zdO$, $t_\z=0$, and $z_2=\zdT$, respectively), 
the resulting pair of coordinate-delay ``points'' will be called an ambiguity pair, see Figure~\ref{fig:sstt_streak}. 
This concept will be helpful in illustrating how delayed scatterers produce range streaks in SAR images (Figure~\ref{fig:Airplane} provides an example).
Consider a point scatterer at $\z=\zd$ that exhibits delayed scattering for $0 \le t_\z \le t_{\max}$. On a standard SAR image $I(0,\y)$, it will show up as a streak in the range direction. 
This streak can be understood as the intersection of the family of ambiguity lines drawn through the support of~$\nu(t_\z,\z)$
\begin{equation*}
	y_1=\zdO
,\quad
	y_2 = \zdT - \frac{c}{\sin\theta}\frac{t_\y-t_\z}{2}
,\quad
	0 \le t_\z \le t_{\max}
	,
\end{equation*}
with the plane $t_\y=0$.
If $t_{\max} \gg B^{-1} $, then the streak extends to the distance of $ct_{\max}/(2\sin\theta) \gg \Delta_\rng$ behind the true location of the scatterer $y_2=\zdT$. 
On a standard SAR image, it can incorrectly be interpreted as a linear instantaneous scatterer between $\zd$ and~$\zd+(ct_{\max}  /(2\sin\theta))\uvector_2$, see Figure~\ref{fig:sstt_streak}. 

On the other hand, the coordinate-delay SAR  provide two values of~$|I|$ for each ambiguity pair (the image $I$ is defined by (\ref{eq:Iconvconv})). The difference between these values is due to the argument~$\psi$ in~\eqref{eq:etazetapsiW}. 
Taking, for definiteness, $t_\z=\tmax$ and using~\eqref{eq:reslnU}--\eqref{eq:twoscales}, we can observe that if 
\begin{equation}
\label{eq:tmax_bPhi}
	\frac{ct_{\max}}{2\sin\theta} \gg \Delta_\una
,\quad\text{or, equivalently,}\quad 
	\frac{\kappa B \tmax}{2} \gg \bPhi,
\end{equation}
then for a delayed scatterer, the value of~$|I|$ taken within the support of~$\nu(t_\z,\z)$ (i.e., at the point~``T'' in Figure~\ref{fig:sstt_streak}) will be larger than that taken in the streak (point~``S''). For an instantaneous scatterer, it will be the other way around.  
Hence, if condition~\eqref{eq:tmax_bPhi} is satisfied, it appears feasible 
to discriminate between these two targets by 
analyzing the ambiguity pairs in the coordinate-delay SAR images.  
Note that for $\tmax \gg 1/B$, this can be realized even in a narrow-aperture case, see~\eqref{eq:kappadef}. 

To be  practical though, the foregoing approach to the detection of  delayed returns  (see also \cite{ferrara-2017}) should be able to deal with extended targets. Scattering off extended targets involves resonant mechanisms that are not captured by merely considering, say, $\nuinst(\z)$ in~\eqref{eq:Istd} with extended support (see \cite{a71e} and \cite[Chapter 7]{sarbook}). Moreover, extended  scatterers exhibit speckle \cite{goodman-76,oliver-98}, a very significant phenomenon in SAR imaging that cannot be simulated by deterministic functions~$\nu(t_\z,\z)$ or~$\nuinst(\z)$. In realistic setups, the effect of speckle combined with high sidelobes of~$\Phi$ has to be addressed. This will be the subject of  subsequent sections. 

\label{text:nusnut}

\section{Speckle in  SAR targets} 
\label{sec:delayedreflectivity}

\subsection{Speckle in homogeneous scatterers} 
\label{sec:ITS}

The scattering properties of radar targets are characterized by multi-scale behavior. On the scale comparable to the wavelength $\lambda = 2\pi c/\omega_0$, the reflectivity is rough. 
It is this small-scale roughness that gives rise to the Bragg resonant mechanism of surface scattering and, in particular, enables the backscattering which is critical for SAR, see \cite{a71e} and \cite[Chapter 7]{sarbook}.

At the same time, the quantities of interest in remote sensing are typically some averaged parameters that characterize the target. They are expected to vary gradually on the scale comparable to the resolution size $\Delta_{\rng,\az}\gg\lambda$. %
In practice, however, this does not happen. The coherent mechanism of SAR imaging leads to the phenomenon of
  speckle. Speckle  makes the image inherit some of the small-scale roughness and thus look ``bumpy'' even on the scale where the parameters of interest are smooth. Speckle is considered a nuisance because it significantly affects our ability to resolve small-scale or low-contrast variations of average reflectivity of the scene. A detailed description of the effect of speckle  can be found in \cite{goodman-84}, \cite{oliver-98}, and~\cite[Chapter~18]{barrett-04}.  

The standard SAR theory for instantaneous targets  is built upon the model of a point target: $\nuinst(\z) =A  \delta(\z-\zd)$, see~\eqref{eq:spacedelta}, where the scatterer location $\zd$ and amplitude $A$ are constants. Despite dominating the SAR literature, neither does this model describe speckle nor can it be easily modified to the case of extended scatterers, i.e., scatterers with non-singular support, see \cite{oliver-98} and~\cite[Chapter~7]{sarbook}. 
The approach described in Section~\ref{sec:ssttW}, see also~\cite{ferrara-2017}, is an extension of the standard SAR to images that depend on one additional ``coordinate'', namely, $t_\y$, see~\eqref{eq:spacetimedelta}. 
As for the speckle, however, it can be described efficiently only in the stochastic framework. 

Next, we are going to build a model for a scatterer with non-singular support in time and/or  space. It will rely on the treatment of extended (sometimes also called distributed) targets as presented in~\cite{oliver-98}. 
In particular, the homogeneous instantaneous reflectivity, or background, is modeled by a random function~$\nuz(t_\z,\z)$ that has the following form (cf.~\eqref{eq:instantaneous}):
\begin{equation}
\label{eq:modelz}
	\nuz(t_\z,\z) = 
	\delta(t_\z)\muz(\z)
	.
\end{equation}
In~\eqref{eq:modelz}, $\muz(\z)$ is a circular Gaussian white random field with the variance~$\sigmasqz$ (see \cite{lapidoth-17,gallager-08a,gallager-08b}):
\begin{equation}
\label{eq:muzasasum}
	\muz(\z) = \myRe\, \muz(\z) + i\myIm\, \muz(\z).
\end{equation}
In (\ref{eq:muzasasum}), $\myRe\,\muz(\z)$  and $\myIm\, \muz(\z)$ are independent real-valued zero-mean white Gaussian noise fields with the intensity  $\frac{\sigmasqz}{2}$:
\begin{equation}
\label{eq:muzReIm}
\begin{aligned}
	\avg{\myRe\,\muz(\z)} = 0
,\quad 
	\big\langle \myRe\,\muz(\z) \myRe\, \muz(\z') \big\rangle 
&=\>
	\frac{\sigmasqz}{2} \delta (\z - \z'),
\\	
	\avg{\myIm\,\muz(\z)} = 0
,\quad 
	\big\langle \myIm\,\muz(\z) \myIm\, \muz(\z') \big\rangle 
&=\>
	\frac{\sigmasqz}{2} \delta (\z - \z'),
\end{aligned}
\end{equation}
so that 
\begin{equation}
\label{eq:muzmuz}
	\avg{\muz(\z)} = 0
,\quad 
	\avg{\muz(\z)\muz(\z')} = 0
,\quad 
	\big\langle\overline{\muz(\z)} \muz(\z') \big\rangle 
=
	\sigmasqz \delta (\z - \z')
	.
\end{equation}
In~\eqref{eq:muzReIm} and~\eqref{eq:muzmuz}, 
$\delta(\z) \equiv \delta(z_1)\delta(z_2)$ according to~\eqref{eq:z3D2D}, 
$\langle\ldots\rangle$ denotes statistical averaging, 
and~$\sigmasqz$ is a deterministic positive  constant that characterizes the statistically averaged reflectivity of the background. 

There is more than one way of defining delta-correlated processes, %
including their non-stationary and multi-dimensional 
versions, 
see \cite{oksendal-03,oksendal-10,lindgren-13,fouque-2007,Papoulis-84,allen-97,lapidoth-17}. %
The notion of a Gaussian white noise as introduced in~\eqref{eq:muzReIm} requires additional clarification. It is a delta-correlated stochastic process with  continuous argument. It is known, however, that random variables with infinite variances cannot be Gaussian. So the word ``Gaussian'' as applied to the delta-correlated process $\muz$ of (\ref{eq:muzasasum})--\eqref{eq:muzmuz} means that it is required to generate a conventional Gaussian process by convolution with the imaging kernel, see~\eqref{eq:Iconvconv}. Moreover, for~$\sigmasqz$ independent of~$\z$, we will call the processes \eqref{eq:muzReIm} and~\eqref{eq:muzmuz} stationary.  
In a more realistic treatment that is not attempted in this study, $\sigmasqz$ can  vary with~$\z$ on the  scale~$\gtrsim \Delta_{\az,\rng}$, see the discussion in the beginning  of this section. 

The use of a random reflectivity function, such as $\nuz$ of~\eqref{eq:modelz}--\eqref{eq:muzmuz}, in the imaging operator~\eqref{eq:Iconvconv} makes the resulting image~${I_\text{b}(t_\y,\y)}$ a random function as well. 
The presence of the delta function in~\eqref{eq:modelz} eliminates integration over~$t_\z$ in~\eqref{eq:Iconvconv}. 
Then, the first two formulas in~\eqref{eq:muzmuz} immediately yield 
\begin{subequations}
\label{eq:avgIzsq}
\begin{equation}
\label{eq:avgI}
	\avg{I_\text{b}(t_\y,\y)} 
= 
	0
,\quad
	\avg{I_\text{b}(t_\y,\y)I_\text{b}(t'_\y,\y')} 
= 
	0
	.
\end{equation}
Multiplying~\eqref{eq:Iconvconv} by its conjugate, taking the average of the result, and using the delta function in~\eqref{eq:muzmuz} along with the explicit form of~$W$ in~\eqref{eq:WviaPhi}, we obtain:
\begin{equation}
\label{eq:the_avgIzsq}
	\avg{|I_\text{b}(t_\y,\y)|^2}
=
	N^2\tau^2\sigmasqz 
	\int \sinc^2(\zeta_0)
	\abssqPhiarg{(y_1-z_1)}{(y_2-z_2)}\,d\z
	,
\end{equation}

\begin{align} 	
\label{eq:Izcov}
	\avg{\overline{I_\text{b}(t_\y,\y)}I_\text{b}(t'_\y,\y')}
=\>&  
	N^2\tau^2\sigmasqz 
	\exp\big({2i\kOtheta(y_2-y'_2)}\big) \exp\big({i\omega_0(t_\y-t'_\y)}\big)
\\ & \nonumber 
\cdot \int \sinc(\zeta_0)
	\overline{\Phiarg{(y_1-z_1)}{(y_2-z_2)}} 
\\ & \nonumber 
\phantom{\cdot \int{}} \sinc(\zeta'_0)
	\Phiarg{(y'_1-z_1)}{(y'_2-z_2)}\,d\z
	,
\end{align}
\end{subequations}
where $d\z=dz_1\,dz_2$, see~\eqref{eq:z3D2D}, and
$$
	\zeta_0 = BT^0\Big|_{t_\z=0} = \frac{B}{\omega_0}\kOtheta(y_2-z_2) + \frac{Bt_\y}{2}
,\quad 
	\zeta'_0 = \frac{B}{\omega_0}\kOtheta(y'_2-z_2) + \frac{Bt'_\y}{2}
	.
$$

As expected for a homogeneous scatterer~\eqref{eq:muzmuz}, expression~\eqref{eq:the_avgIzsq} does not depend on $\y$ because the  integrand on the right-hand side of \eqref{eq:the_avgIzsq} depends on the integration variable $\z$ only via~$(\y-\z)$, and the integration is performed over the entire space. 
We can see that the values of~$I$ become decorrelated:
\begin{equation}
\label{eq:decorr_ll}
	\big|\big\langle \overline{I_\text{b}(t_\y,\y)}I_\text{b}(t'_\y,\y') \big\rangle \big|^2
	\ll
	{\big\langle |I_\text{b}(t_\y,\y)|^2\big\rangle \big\langle |I_\text{b}(t'_\y,\y')|^2\big\rangle} 
	,
\end{equation}
when the peaks of the functions in the integrand of~\eqref{eq:Izcov} significantly separate:
\begin{equation}
\label{eq:decorrIz}
	|\zeta_0-\zeta'_0| \gtrsim \pi
\quad\text{or}\quad 
	|\eta-\eta'| \equiv 
	\kOtheta\phiT|y_1-y'_1|\gtrsim \max(\pi, \pi \kappa/2).
\end{equation}
Note that the second inequality in~\eqref{eq:decorrIz} addresses both the first and second argument of~$\Phi$, 
see \eqref{eq:ggDeltaAz} and the discussion right before it.
On the other hand, if we fix two coordinate-delay ``points,'' $(t_\y,\y)$ and~$(t'_\y,\y')$,  on the same ambiguity line: 
$$
	y_1=y'_1
,\quad
	 \frac{B}{\omega_0}\kOtheta y_2  + \frac{Bt_\y}{2} = 
	 \frac{B}{\omega_0}\kOtheta y'_2+ \frac{Bt'_\y}{2}
,\quad
	|t_\y-t'_\y| = \text{const}
	,
$$
e.g., take an ambiguity pair as shown in Figure~\ref{fig:sstt_streak}, then 
the inequality in~\eqref{eq:decorr_ll} becomes an equality 
as $\kappa\to 0$. 
In other words, in the narrow-aperture case, the image values along the ambiguity lines are strongly correlated.

The Gaussianity of the random field~$I_\text{b}(t_\y,\y)$ is corroborated by  experimental evidence~\cite{ulaby-89,oliver-98,goodman-84}.  In our analysis, it is a requirement imposed on $\muz(\z)$, see the discussion after \eqref{eq:muzmuz}. As a consequence, we have: 
\begin{equation}
\label{eq:specklemoments}
	\Var\big(|I_\text{b}(t_\y,\y)|^2\big) = \avg{|I_\text{b}(t_\y,\y)|^4}-\avg{|I_\text{b}(t_\y,\y)|^2}^2=\avg{|I_\text{b}(t_\y,\y)|^2}^2
	.
\end{equation}
It means, in particular, that 
when the image is a circular Gaussian random field, the difference between the intensity $|I_\text{b}|^2$ in the neighboring pixels may often be comparable to its average. 
It is this property of the random field $I_\text{b}(0,\y)$ that is responsible for speckle \cite{oliver-98,goodman-76,goodman-84}; it creates visual roughness and presents major difficulties in analyzing SAR images of extended  scatterers. 

The derivation of relations \eqref{eq:modelz}--\eqref{eq:muzmuz} from  first principles, based on the shape and dielectric properties of the scatterer, may be complicated. The model \eqref{eq:modelz}--\eqref{eq:muzmuz} is still convenient, because when combined with the assumption of circular Gaussianity,  it allows us to establish certain useful relations between the image moments, see \eqref{eq:avgIzsq} and~\eqref{eq:specklemoments}. The latter, in turn,  facilitate the efficient image processing.

\subsection{Inhomogeneous image components}
\label{sec:nusnut}

\begin{figure}[ht]
\begin{center}
\hspace{-1.3cm}
	\includegraphics[width=3in]{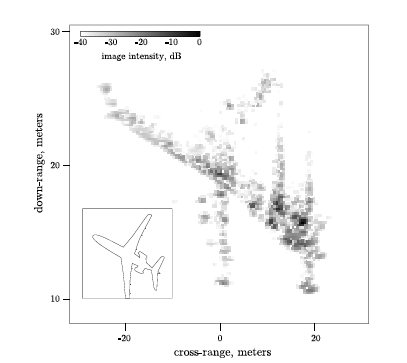}
\vspace{-0.2cm}
\end{center}
\caption{This image of an aircraft shows streaks due to engine inlets \cite{borden-98}.}

\label{fig:Airplane}
\end{figure}

In SAR images, the streaks due to delayed scattering (see, e.g., \cite[Figure~1]{ferrara-2017},  \cite{trintinalia-97}, as well as Figure~\ref{fig:Airplane})%
\footnote{
The examples where streaks can be clearly seen are usually found in Inverse SAR (ISAR) observations of aircraft where there is no background obscuring the streak. An example where a streak is visible on a SAR image of the Earth surface can be found in~\cite[Figure~1]{ferrara-2017}. 
}
appear rugged, similarly to the speckle that was considered in Section~\ref{sec:ITS}.  
For this reason, we will introduce two extensions of the speckle-producing scatterer model~\eqref{eq:modelz}--\eqref{eq:muzmuz}. 
First, we will describe a scatterer localized in space and exhibiting a delta-correlated delayed return. Henceforth,  this model will  be called a ``t-scatterer:'' %
\begin{equation}
\label{eq:modeltz}
	\nut(t_\z,\z) \equiv \nut(t_\z,\z; \zd) =  
	\mut(t_\z)\delta(\z-\zd)
	.
\end{equation}
In (\ref{eq:modeltz}), $\mut$ is a non-stationary circular Gaussian white noise (c.f.\ \eqref{eq:muzasasum}--(\ref{eq:muzmuz})):
\begin{equation}
\label{eq:mutmut}
	\avg{\mut(t)} = 0
,\quad
	\big\langle \mut(t) \mut(t') \big\rangle = 0
,\quad
	\big\langle\overline{\mut(t)} \mut(t') \big\rangle 
	= 
	\sigmasqt F_t(Bt/2) \delta (t-t') 
	, 
\end{equation} 
where $\zd$ is the location of the scatterer. 
Similarly to~\eqref{eq:muzmuz}, we will call $\sigmasqt$ the reflectivity of the t-scatterer, whereas $F_t=F_t(\zeta)$ defines the dimensionless  intensity of the return as  a function of time.  
To make sure that the scattering model (\ref{eq:mutmut}) is causal, we require that $F_t(Bt/2) = 0$ for $t<0$, see~\eqref{eq:linscattering}. For  $t>0$, we 
 assume that $F_t(Bt/2)\geqslant 0$. In~\cite{allen-97}, the existence of a process that satisfies~\eqref{eq:mutmut} is shown when the function $F_t(\zeta)$ is integrable on $(-\infty,\infty)$. From the standpoint of physics, this also limits the total power reflected by a scatterer of this type. 
In this work, we will take $F_t$ as an indicator function: 
\begin{equation}
\label{eq:Heaviside}
	F_t(\zeta) = \begin{cases} 1, & 0\leqslant \zeta\leqslant\zeta_{\max},\\ 0, & \text{otherwise,} \end{cases}
\end{equation}
where the value of $\zeta_{\max}$ will be discussed later (see Section~\ref{sec:imagesInhom}). 
Then, similarly to~$I_\text{b}$, the image~$I_t$, which is given by \eqref{eq:Iconvconv} applied to $\nut$ of (\ref{eq:modeltz}), is circular Gaussian, and the properties~\eqref{eq:avgI} and~\eqref{eq:specklemoments} hold for $I_t(t_\y,\y)$ as well. 

In standard SAR images, a t-scatterer described by~\eqref{eq:modeltz}--\eqref{eq:mutmut} may be confused with a linearly shaped inhomogeneity of the background intensity  aligned with the range direction (see Figure~\ref{fig:sstt_streak}). The corresponding instantaneous scatterer will henceforth be called  an ``s-scatterer:''
\begin{equation}
\label{eq:modelsz}
	\nus(t_\z,\z) \equiv \nus(t_\z,\z; \zd) = 
	\delta(t_\z) \delta(z_1-\zdO) \mus(z_2-\zdT),
\end{equation}
where 
\begin{equation}
\label{eq:musmus}
	\avg{\mus(s)} = 0
,\quad
	\big\langle \mus(s) \mus(s') \big\rangle = 0
,\quad
	\big\langle\overline{\mus(s)} \mus(s') \big\rangle
=
	\sigmasqs F_s(B\kOtheta s / \omega_0) \delta (s-s') 	
	.
\end{equation} 
Linearly-shaped inhomogeneities of instantaneous reflectivity, as in \eqref{eq:modelsz}--\eqref{eq:musmus}, may be representatives of roads, fences, edges of buildings, pipelines, etc. 
The three factors on the right hand side of the last equation in~\eqref{eq:musmus} are similar to those in~\eqref{eq:mutmut}. The process $\mus$ is a non-stationary circular Gaussian white noise.
Causality imposes no restriction on~$F_s$. However, similarly to~$\nut$ of \eqref{eq:modeltz}--\eqref{eq:mutmut}, we assume that  $F_s(\zeta)=0$ for $\zeta < 0$.

We have introduced the scatterer models \eqref{eq:modeltz}--\eqref{eq:mutmut} and \eqref{eq:modelsz}--\eqref{eq:musmus}  as an alternative to the  space-time point scatterer~\eqref{eq:spacetimedelta} studied  in~\cite{ferrara-2017}. In particular, the form \eqref{eq:modeltz}--\eqref{eq:mutmut} allows us to gather the signals reflected from the same target with different delays and analyze the resulting data. 
Similarly to the models developed in~\cite[Chapter~4]{chen-2002b} and~\cite{borden-98}, we will only analyze  the scenarios where the response delay of the scatterer~$\nut$ of~\eqref{eq:modeltz}--\eqref{eq:mutmut} does not exceed a certain predetermined maximum. However, the correlation properties of the scattered signal due to $\nut$ will differ from those considered in \cite{chen-2002b} or \cite{borden-98}.

\section{Discrimination between scatterer types in the presence of background and noise}
\label{sec:discrimination}

\subsection{Images due to inhomogeneous targets} 
\label{sec:imagesInhom}

Let us introduce the dimensionless coordinates for the image, see~\eqref{eq:etazetapsi}, with the origin at $(t_\y,\y)=(0, \zd)$: 
\begin{equation}
\label{eq:etazetapsid}
\begin{aligned}
	(\eta_\y,\zetay,\psi_\y) = (\eta,\zeta,\psi)\Big|_{t_\z = 0, \; \z=\zd}
	.
\end{aligned}
\end{equation}
Similarly to~\eqref{eq:avgIzsq}, it is easy to obtain the expected values of image intensities for the scatterers described in Section~\ref{sec:nusnut}. 
The ambiguity lines in the coordinates~\eqref{eq:etazetapsid} are  $\big\{\eta_\y = \text{const}, \; \zeta_\y=\text{const}\big\}$. 
Substituting \eqref{eq:modeltz}--\eqref{eq:musmus} into~\eqref{eq:Iconvconv} and using \eqref{eq:WviaPhi}, we have:
\begin{align}
\nonumber
	\avg{|I_s|^2}(\eta_\y,\zetay,\psi_\y)
&=\>
	N^2\tau^2\sigmasqs 	\frac{\omega_0}{B\kOtheta}
\\\nonumber & \phantom{=\>} \cdot 
	\int_0^\infty F_s(\zeta') \sinc^2(\zetay-\zeta') 
	\Big|\Phi\Big[\eta_\y,\kappa\Big(\frac{\zetay+\psi_\y}{2}-\zeta'\Big)\Big]\Big|^2\,d\zeta'
	,
\\ \nonumber
	\avg{|I_t|^2}(\eta_\y,\zetay,\psi_\y)
&=\>
	N^2\tau^2\sigmasqt 	\frac{2}{B}
	\Big|\Phi\Big[\eta_\y,\kappa\frac{\zetay+\psi_\y}{2}\Big] \Big|^2
\\ & \phantom{=\>} \cdot 	
	\int_0^\infty F_t(\zeta') \sinc^2(\zetay-\zeta') 
	\,d\zeta'
	.
	\label{eq:IsIt}
\end{align}
Two different expressions \eqref{eq:IsIt} yield two different locations of the maximum   image intensity (average) along the ambiguity line. 
Consider, for simplicity, the case of $\eta_\y = 0$ and both $F_s(\zeta)$ and $F_t(\zeta)$~given by~\eqref{eq:Heaviside}. Note that, $ \sinc^2(\zetay-\zeta') $ has its central lobe on the interval $[-\pi,\pi]$ of its argument and decays quadratically for large arguments. 
Take $\zetamax \gg \pi$. Then, for $\zetay$ satisfying $\pi \ll \zetay \ll \zetamax$, see~\eqref{eq:Heaviside}, 
the value of either integral in (\ref{eq:IsIt}) will not change significantly if the integration limits are replaced with $(-\infty,\infty)$ and $F_t(\zeta')$ and $F_s(\zeta')$  are replaced with 1 for all arguments.
\begin{align}
\nonumber
	\avg{|I_s|^2}(\eta_\y,\zetay,\psi_\y)
&\approx\>
	N^2\tau^2\sigmasqs 	\frac{\omega_0}{B\kOtheta}
\\\nonumber & \phantom{=\>} \cdot 
	\int_{-\infty}^\infty  \sinc^2(\zetay-\zeta') 
	\Big|\Phi\Big[\eta_\y,\kappa\Big(\frac{\zetay+\psi_\y}{2}-\zeta'\Big)\Big]\Big|^2\,d\zeta'
	,
\\ \nonumber
	\avg{|I_t|^2}(\eta_\y,\zetay,\psi_\y)
&\approx\>
	N^2\tau^2\sigmasqt 	\frac{2}{B}
	\Big|\Phi\Big[\eta_\y,\kappa\frac{\zetay+\psi_\y}{2}\Big] \Big|^2
\\ & \phantom{=\>} \cdot 	
	\int_{-\infty}^\infty \sinc^2(\zetay-\zeta') 
	\,d\zeta'
	.
	\label{eq:IsIt2}
\end{align}

Therefore, the maximum of $\avg{|I_s|^2}$ as a function of $\psi_\y$ is achieved when the peaks of $\sinc^2(\zetay-\zeta')$ and $|\Phi
\big(0,\kappa(\ldots - \zeta')\big)|^2$ as functions of~$\zeta'$ overlap, i.e., for $\psi_\y \approx \zetay$.%
\footnote{Since functions $\sinc^2(\cdot)$ and $|\Phi|^2(0,\cdot)$ are even and have local maxima at zero, it can be shown that $\zeta_\y$ is a local maximum of $\avg{|I_s|^2}(0,\zetay,\cdot)$ given by~\eqref{eq:IsIt}. 
We can also demonstrate numerically 
that it is a global maximum. 
}
As for $\avg{|I_t|^2}$ as a function of  $\psi_\y$, it reaches its maximum value when $|\Phi|^2$ peaks with respect to its second argument, i.e., at  $\psi_\y = -\zetay$.
For a given ambiguity line, the locations of these maxima correspond to the points ``S'' and ``T'', respectively, of the ambiguity pair, see Figure~\ref{fig:sstt_streak}. Thus, one can see a  similarity between the expectations~\eqref{eq:IsIt2} and images of a deterministic point target~\eqref{eq:spacetimedelta} considered in Section~\ref{sec:Wproperties}. 

The main lobe of~$|\Phi(0,\cdot)|^2$ can be thought of as confined to the interval $[-\bPhi,\bPhi]$,
 see the right panel in Figure~\ref{fig:Phimarginals}. Then, we may argue that for a given ambiguity line  the peaks of $\avg{|I_s(0,\zetay,\psi_\y)|^2}$ and~$\avg{|I_t(0,\zetay,\psi_\y)|^2}$ are well separated if 
$$
	\kappa\zetay \gtrsim \bPhi, 
$$ 
which is similar to condition~\eqref{eq:tmax_bPhi} obtained for the deterministic case. 
It is also easy to see that the dependence of both expressions \eqref{eq:IsIt2} on~$\psi_\y$ becomes weaker and eventually vanishes as $\kappa\to 0$. 

We have just shown that for a fixed~$\kappa>0$, the bigger the $\zeta_\y$, the better the separation between the peaks of~$\avg{|I_s(0,\zeta_\y,\cdot,\psi_\y)|^2}$ and $\avg{|I_t(0,\zeta_\y,\psi_\y)|^2}$. 
Hence, for a discrimination procedure outlined after \eqref{eq:tmax_bPhi}, we should take the largest possible values of~$\zeta_\y$. However, the functions $F_s$ and~$F_t$ differ from zero only on a finite interval, see (\ref{eq:Heaviside}), and there are several considerations that lead to choosing a particular value of $\zeta_{\max}$.
 On one hand, we may have an a priori knowledge about the maximum duration $t_{\max}$ of the delayed response, so that
\begin{subequations}
\label{eq:zetamax}
\begin{equation}
\label{eq:tmax}
	\zeta_\y \le \zetamax \le \frac{Bt_{\max}}{2}.
\end{equation}
On the other hand, there may exist a maximum extent in range, $s_{\max}$, at which the background~\eqref{eq:muzmuz} can be considered homogeneous. Hence, 
\begin{equation}
\label{eq:smax}
	\zeta_\y \le \zetamax \le \frac{B\kOtheta s_{\max}}{\omega_0}.
\end{equation}
\end{subequations}
Hereafter, we will consider $\zetamax$ to be a  parameter of the formulation that satisfies inequalities~\eqref{eq:zetamax}. Moreover,  given that $\zeta_{\max}\geqslant\zetay+\text{const}$,
we must have $\kappa \zetamax \gtrsim \bPhi$ in order to distinguish between the statistically averaged image intensities due to delayed and instantaneous targets as defined in Section~\ref{sec:nusnut}.

Recalling the definition of $\kappa$ in~\eqref{eq:kappadef},  we can see that the inequality $\kappa \zetamax \gtrsim \bPhi$  yields a condition that the angle $\phiT$ should satisfy:
\begin{equation}
\label{eq:kappazetamax}
	\bPhi \lesssim \kappa \zetamax = \phiT^2 \cdot \min\big( \omega_0t_{\max}/2, \kOtheta s_{\max}\big)
	.
\end{equation}
The following geometrical interpretation is therefore possible. In the target domain, the space-time point $(t_{\max}, \zd)$ is ambiguous with the ``instantaneous'' point $(0,\zd +s\uvector_2)$, where $s=ct_{\max}/(2\sin\theta)$. The condition $\kappa\zetamax\gtrsim \bPhi$ means that the two-way travel distance difference $2\big(|\x-\zd| - |\x-(\zd+s\uvector_2)|\big)$ varies by at least $ (\bPhi/2\pi)\lambda \approx 3\lambda$ as $\x$ scans the synthetic aperture of angular width~$\phiT$. An analogy can be found in the expression~\eqref{eq:DeltaRA} for~$\Delta_\az$: it corresponds to the angle~$\phiT$ such that for a pair of points $\z$ and $(\z+\Delta_\az \uvector_1)$, the similar variation of the two-way travel path difference~$2c\Tlinphi$ (see~\eqref{eq:T0}) equals to~$\lambda$.

\subsection{Image sampling and the approach to discrimination in general}
\label{sec:discriminationIdea}

The effect of speckle on the detection of maxima of the quantities  \eqref{eq:IsIt2} 
will be similar to that of noise. 
A well-known  strategy for detecting a weak signal in the presence of noise is to collect the signal over a sufficiently long time so that the effect of noise would  average out and hence decrease due to its statistical properties. We will adjust this strategy to the case  of delayed responses in~SAR imaging.  The ``weak signal'' will be  the variation of the average image intensity along the ambiguity lines, 
see Figure~\ref{fig:sstt_streak} and the discussion around equations~\eqref{eq:IsIt}--\eqref{eq:IsIt2}. 
As a counterpart to 
collecting the weak signal over a long interval, we will sample the image at multiple spatial locations and delays.

We assume that we have to distinguish between two possible configurations of scatterers in the target: 
\begin{subequations}
\label{eq:6nu}
\begin{equation}
\label{eq:6nus}
	\nu(t_\z,\z) = 
	\nu_\text{s-model}(t_\z,\z;\zd) = \nuz(t_\z,\z)+\nus(t_\z,\z;\zd)
\end{equation}
and  
\begin{equation}
\label{eq:6nut}
	\nu(t_\z,\z) = 
	\nu_\text{t-model}(t_\z,\z;\zd) = \nuz(t_\z,\z)+\nut(t_\z,\z;\zd)
	,
\end{equation}
\end{subequations}
where $\nuz$, $\nut$, and $\nus$ are defined in~\eqref{eq:modelz}, \eqref{eq:modeltz}, and~\eqref{eq:modelsz}, respectively, and~$\zd$ in \eqref{eq:6nus} and~\eqref{eq:6nut} is the same. The names ``s-model'' and ``t-model'' are intended to match the terms ``s-scatterer'' and ``t-scatterer'' introduced in Section~\ref{sec:nusnut}, see~\eqref{eq:modeltz} and~\eqref{eq:modelsz}. Accordingly, the corresponding total images are  given by either  
\begin{subequations}
\label{eq:16Ity}
\begin{equation}
\label{eq:16Itys}
	I_\text{s-model}(t,\y;\zd) = I_\text{b}(t,\y) + I_\text{n}(t,\y)  + I_s(t,\y;\zd) 
	\phantom{.}
\end{equation}
or 
\begin{equation}
\label{eq:16Ityt}
	I_\text{t-model}(t,\y;\zd) = I_\text{b}(t,\y)  + I_\text{n}(t,\y)+ I_t(t,\y;\zd)
	.
\end{equation}
\end{subequations}
In formulae \eqref{eq:16Ity}, we have introduced the terms $I_\text{n}$ to represent the  receiver noise and processing errors. The  noise $I_\text{n}$ is taken as a circular Gaussian process with independent  samples.
Note that the background $I_\text{b}$ is also a circular Gaussian process, but its correlation properties are different --- they are determined by the imaging kernel (\ref{eq:WviaPhi}), see Section~\ref{sec:ITS}. 
A detailed definition of the noise term $I_\text{n}$ is given in \ref{sec:FGH}. 

Our discrimination procedure will be based on the  analysis of the image $I(t_\y,\y)$. In addition to the image per se, we assume that  some a priori information is known about the target, such as the reflectivity profiles $F_{s,t}(\zeta)$. We will also assume that we know~$\zd$. We can find candidate locations for $\zd$ as the locations of sharp increases in the intensity of a standard  SAR  image $I(0, \y)$ in the range direction. The latter can potentially be identified using edge detection \cite{canny-86,basu-02,ziou-98,marr-80}, a technique that we do not discuss in the current paper.

As long as $\zd$ is known, we can define the ambiguity lines for  inhomogeneous scatterers, see Figure~\ref{fig:sstt_streak}. To tell between the scatterer types, we would ideally want to obtain a large amount of data. However, the 
samples of an image given by convolution~\eqref{eq:Iconvconv} will not be independent, with the correlation determined by the kernel~$W(t_\y,\y; t_\z,\z)$ of~\eqref{eq:WviaPhi}. The geometry of the central peak and sidelobes of~$W$ has been analyzed in Section~\ref{sec:Wproperties}. From this analysis, we derive  that the autocorrelation of~$I(t_\y,\y)$ quickly decreases across the ambiguity directions (Section~\ref{sec:ITS}). For this reason, we will take a finite number of values of~$\zetay$ according to 
\begin{equation}
\label{eq:zetaym}
	\zetaym = \pi m, \quad \eta_m=0
,\quad\text{where}\quad	
	m\in\mathbb{N}
,\quad	
	\zetamin < \zetaym \le \zetamax
	.
\end{equation}	
In~\eqref{eq:zetaym}, $\zetamax$ is a  parameter of the formulation, see also~\eqref{eq:zetamax}. In addition, we have  introduced another parameter, $\zetamin \gtrsim \pi$, to cut off the transitional effects due to the behavior   of~$F_s(\zeta)$ and~$F_t(\zeta)$ given by~\eqref{eq:Heaviside} in the vicinity of~$\zeta = 0$. 
The set of locations $\{\zetaym\}$ given by \eqref{eq:zetaym} defines a family of ambiguity lines via~\eqref{eq:etazetapsid} such that we can treat the samples of the image taken on different ambiguity lines as independent. 

Still,  to resolve the range-delay ambiguity, we should take more than one measurement of~$I$ on each ambiguity line. In doing so, we cannot avoid dealing with strongly correlated image samples, because the autocorrelation of~$I(t_\y,\y)$ decreases slowly along the ambiguity direction (Section~\ref{sec:ITS}). In this work, for each  ambiguity line introduced for $(\zetaym,\eta_m)$ of (\ref{eq:zetaym}), we will take  two values of $I$ that correspond to the ambiguity pair ``S'' and~``T'', see Figure~\ref{fig:sstt_streak}. In other words, we will consider  a set of pairs~$\big(\IS_\streak(\zetaym) ,\IT_\streak(\zetaym)\big)$:
\begin{equation}
\label{eq:streakST}
\begin{aligned}
	\IS_{\streak,m}
&\bydef\>
	I(t_\y,\y)\Big|_{t_\y=0,\; \y = \zd + \uvector_2  \omega_0\zetaym/(B\kOtheta)}
	\;,
\\ 
	\IT_{\streak,m}
&\bydef\> 
	I(t_\y,\y)\Big|_{t_\y= 2\zetaym/B,\; \y = \zd}	
	\;,
\end{aligned}
\end{equation}
where~$\zetaym$ is given by~\eqref{eq:zetaym}. 
Choosing the  locations ``S'' and~``T'' on a given ambiguity line has the advantage of maximizing the expectation of the intensity of at least one of the two possible inhomogeneous images, $|I_s|^2$ or $|I_t|^2$. This is beneficial in the presence of  fluctuations due to the background and noise. 

\newcommand{\homog}{\text{hom}}
\newcommand{\yhomm}{\y_k}
\newcommand{\yhommp}{\y_{k'}}

In addition to the samples~\eqref{eq:streakST} taken at the streak, we would like to see whether sampling the homogeneous part of the image around the streak:
\begin{equation}
\label{eq:16Ityb}
	I(t_\y,\y) = I_\text{b}(t_\y,\y) + I_\text{n}(t_\y,\y)	
\end{equation}
may affect the performance of the discrimination algorithm to be built.
To that end, we 
take a set of points~$\{\yhomm\}$ such that the image values at  $(t_\y,\y)=(0, \yhomm)$ have low correlation with each other and with the streak samples, 
see~\eqref{eq:decorrIz}, 
and consider the following homogeneous samples:
\begin{equation}
\label{eq:homST}
\begin{aligned}
	\IS_{\homog,k} 
&\bydef\>
	I(t_\y,\y)\Big|_{t_\y=0,\; \y = \yhomm}
	\;,
\\ 
	\IT_{\homog,k} 
&\bydef\> 
	I(t_\y,\y)\Big|_{t_\y= 2\zetamax/B,\; \y = \yhomm - \uvector_2 \omega_0\zetamax/(B\kOtheta)}	
	.
\end{aligned}
\end{equation}
Their  effect  on the discrimination quality is outlined in Section~\ref{sec:ssttResults}. %

\subsection{Anticipated statistics of the sampled image}
\label{sec:anticipatedStatistics}

The second moments of the various image components in~\eqref{eq:16Ity} can be obtained by substituting the scatterer models $\nuz$, $\nut$, and $\nus$ (see formulae \eqref{eq:muzmuz}, \eqref{eq:mutmut}, and~\eqref{eq:musmus}, respectively) into the imaging operator~\eqref{eq:Iconvconv} with the kernel~\eqref{eq:WviaPhi}. The details of the calculations are given in \ref{sec:FGH}.  The resulting expressions have the following form (cf.~\eqref{eq:avgIzsq} and~\eqref{eq:IsIt}): 
\newcommand{\operF}[2]{\,#1[#2]\,}
\begin{equation}
\label{eq:3avg}
\begin{aligned}
	\avg{|\ISalpha(\zeta)|^2} 
&=\>
	\sigmasqalpha K_\alpha \operF{G^S_\alpha}{F_\alpha}(\zeta) 
	,
\\ 
	\avg{|\ITalpha(\zeta)|^2} 
&=\>
	\sigmasqalpha K_\alpha \operF{G^T_\alpha}{F_\alpha}(\zeta)
	,
\\
	\avg{\roverline{\ITalpha(\zeta)} \ISalpha(\zeta)} 
&=\>
	\sigmasqalpha K_\alpha \operF{H_\alpha}{F_\alpha}(\zeta)	
	.
\end{aligned}
\end{equation}
In~\eqref{eq:3avg}, we use the following notations:
\begin{itemize}
\item
The superscripts $S$ and $T$ are the same as introduced in~\eqref{eq:streakST}; they refer to the components of an ambiguity pair, see Figure~\ref{fig:sstt_streak}. A certain fixed value of~$\zd$ is always assumed. 

\item
The subscripts $\alpha \in \{\text{b},t,s\}$ refer to the scatterer models $\nuz$, $\nut$, and $\nus$, respectively, whereas $\sigma^2_{\text{b},t,s}$ are the corresponding scattering intensities. 

\item 
The operators $G^S_{\text{b},s,t}$, $G^T_{\text{b},s,t}$, and $H_{\text{b},s,t}$ act on the functions $F_{\text{b},s,t}$. These operators, as well as  the scalars $K_{\text{b},s,t}$, are defined in \ref{sec:FGH}. 

\item
The case~$\alpha=\text{n}$ corresponds to the noise term~$I_\text{n}$ introduced in \eqref{eq:16Ity} and~\eqref{eq:16Ityb}. The 
corresponding  intensity $\sigma^2_\text{n}$, constant $K_\text{n}$, function~$F_\text{n}(\zeta)$, and operators 
$\operF{G^{S,T}_\text{n}}{F_n}$ and $\operF{H_\text{n}}{F_n}$ are also defined in \ref{sec:FGH}.

\end{itemize}

Using~\eqref{eq:3avg}, we express the statistics of the image samples defined by~\eqref{eq:streakST} for the case of an instantaneous inhomogeneous target~\eqref{eq:6nus} as follows: 
\begin{equation}
\label{eq:streakISTalpha_s}
	\text{s-model:}\quad 
	\begin{aligned}
		\vphantom{\bigg\langle}
		\avg{|I^{S,T}_{\streak,m}|^2}
	&=\> 
		\sum_{\alpha \in \{\text{b},\text{n},s\}} 
		\sigmasqalpha K_\alpha \operF{G^{S,T}_\alpha}{F_\alpha}(\zetaym) 
		,
	\\
		\avg{\overline{\IT_{\streak,m}} \IS_{\streak,m}}
	&=\> 
		\sum_{\alpha \in \{\text{b},\text{n},s\}} 
		\sigmasqalpha K_\alpha \operF{H_\alpha}{F_\alpha}(\zetaym)
	.
	\end{aligned}
\end{equation}
For the homogeneous samples \eqref{eq:homST} %
we have:
\begin{equation}
\label{eq:backgrISTalpha}
	\text{s-model or t-model:}\quad 	
	\begin{aligned}
		\vphantom{\bigg\langle}
		\avg{|I^{S,T}_{\homog,k}|^2}
	&=\> 
		\sum_{\alpha \in \{\text{b},\text{n}\}}
		\sigmasqalpha K_\alpha \operF{G^{S,T}_\alpha}{F_\alpha}(\zetamax),
\\
		\avg{\overline{\IT_{\homog,k}} \IS_{\homog,k}}
	&=\> 
		\sum_{\alpha \in \{\text{b},\text{n}\}}
		\sigmasqalpha K_\alpha \operF{H_\alpha}{F_\alpha}(\zetamax)
	\end{aligned}
\end{equation}
For the delayed target~\eqref{eq:6nut}, the summation over $\alpha \in \{\text{b},\text{n}, s\}$ on the right-hand side of formulae \eqref{eq:streakISTalpha_s} is replaced with the summation over $\alpha \in \{\text{b},\text{n}, t\}$:
\begin{equation}
\label{eq:streakISTalpha_t}
	\text{t-model:}\quad 
	\begin{aligned}
		\vphantom{\bigg\langle}
		\avg{|I^{S,T}_{\streak,m}|^2}
	&=\> 
		\sum_{\alpha \in \{\text{b},\text{n},t\}} 
		\sigmasqalpha K_\alpha \operF{G^{S,T}_\alpha}{F_\alpha}(\zetaym) 
		,
	\\
		\avg{\overline{\IT_{\streak,m}} \IS_{\streak,m}}
	&=\> 
		\sum_{\alpha \in \{\text{b},\text{n},t\}} 
		\sigmasqalpha K_\alpha \operF{H_\alpha}{F_\alpha}(\zetaym)
	.
	\end{aligned}
\end{equation}

\subsection{Description of the discrimination procedure} 
\label{sec:st_discrimination}

The problem of discrimination between the scenarios~\eqref{eq:6nus} and~\eqref{eq:6nut} can now be formulated as follows. For a given image, our dataset consists of two sets of pairs of complex numbers. There are streak pairs $\big(\IS_{\streak,m},\IT_{\streak,m}\big)$ given by (\ref{eq:streakST}),  where $1 \le m \le N_\streak$ and $N_\streak$ is determined by~\eqref{eq:zetaym}. In addition, there are   homogeneous pairs~$(\IS_{\homog,k},\IT_{\homog,k})$ given by (\ref{eq:homST}), where $1 \le k \le N_\homog$ and $N_\homog$ is a parameter of the formulation.
We also have two mathematical models for the statistics of these pairs. The first model is given by \eqref{eq:streakISTalpha_s}--\eqref{eq:backgrISTalpha}, whereas the second model is given by \eqref{eq:backgrISTalpha}--\eqref{eq:streakISTalpha_t}. These models contain the unknown parameters $\sigmasqs$ and $\sigmasqt$, respectively. Moreover, they share two common unknowns $\sigmasqz$ and~$\sigmasqn$. The problem of discrimination is to choose the model that fits the given dataset better than the other model does.

 For each of the two models, our discrimination algorithm will seek the set of unknowns~$\sigmasqalpha$ that maximizes the probability density of the dataset. Then, we will choose the model that yields the larger of the two maxima. 
This is the same idea as that behind the maximum likelihood (ML) approach \cite{Mendenhall-73,oliver-98}.

For a given scatterer type $\alpha$, let~$\mathbf{r}$ be a vector  of four real  Gaussian random variables that define the quantities  on the left-hand side of~\eqref{eq:3avg}:
\newcommand{\TT}{{\text{{\bf T}}}}
\begin{equation}
\label{eq:vecrdef}
	\mathbf{r}  \bydef \veclist{\myRe\,\ISalpha(\zeta), \myIm\,\ISalpha(\zeta),\myRe\,\ITalpha(\zeta), \myIm\,\ITalpha(\zeta) }^\TT
	.
\end{equation} 
Introduce the following brief notations for the right-hand sides of~\eqref{eq:3avg}: 
\begin{equation}
\label{eq:ABCDdef}
\begin{aligned}
	A &= \sigmasqalpha K_\alpha \operF{G^S_\alpha}{F_\alpha}(\zeta) 
,\quad&
	C &= \myRe\big(\sigmasqalpha K_\alpha \operF{H_\alpha}{F_\alpha}(\zeta)	\big)
, \\
	B &= \sigmasqalpha K_\alpha \operF{G^T_\alpha}{F_\alpha}(\zeta)	
,\quad&
	D &= \myIm\big(\sigmasqalpha K_\alpha \operF{H_\alpha}{F_\alpha}(\zeta)	\big) 
	,
\end{aligned}
\end{equation}
so that 
\begin{equation}
\label{eq:alphaABCD}
\begin{aligned}
	\avg{|\ISalpha(\zeta)|^2} 
=%
	A
,\quad %
	\avg{|\ITalpha(\zeta)|^2} 
=%
	B
,\quad %
	\avg{\roverline{\ITalpha(\zeta)} \ISalpha(\zeta)} 
=%
	C+iD	
	.
\end{aligned}
\end{equation}
We have $A,B,C,D\in\mathbb{R}$ and $A,B>0$. Recall that $I^S_\alpha$ and~$I^T_\alpha$ are circular Gaussian variables satisfying $\avg{\ITalpha(\zeta)\ISalpha(\zeta)} = 0$, see~\eqref{eq:avgI}. Then, we have 
\begin{equation}
\label{eq:MABCD}
	\avg{\mathbf{r}\mathbf{r}^\TT} = \frac{1}{2}
	\begin{pmatrix}
		A & 0 & C & -D \\
		0 & A & D & C \\
		C & D & B & 0 \\
		-D & C & 0 & B
	\end{pmatrix} 
	\bydef \mathbf{M}
	.
\end{equation}
The condition that the matrix~$\mathbf{M}$ is positive semidefinite translates into~$AB \ge C^2+D^2$, or 
$ \operF{G^S_\alpha}{F_\alpha}(\zeta) \cdot \operF{G^T_\alpha}{F_\alpha}(\zeta)  \ge  |\operF{H_\alpha}{F_\alpha}(\zeta) |^2$. The latter is a consequence of the Schwartz inequality applied to~\eqref{eq:alphaABCD}. 

Next, we switch to the extended notation by adding the indices~$\alpha$ and~$j$ to~$\mathbf{r}$, $A$, $B$, $C$, $D$, and~$\mathbf{M}$: 
$$
	 (\mathbf{r},  A,  B,  C,  D, \mathbf{M}) \mapsto 
	 (\mathbf{r}_{\alpha,j} ,  A_{\alpha,j} ,  B_{\alpha,j},  C_{\alpha,j},  D_{\alpha,j}, \mathbf{M}_{\alpha,j}) 
	 ,
$$
where~$\alpha$ denotes the scatterer type  as in~\eqref{eq:3avg}, and $j$ indexes all pairs of samples in the dataset going first through the streak samples~\eqref{eq:streakST} and then through the  homogeneous samples~\eqref{eq:homST}, so that $1\le j \le (N_\streak+N_\homog)$. In addition,  define $\zeta_j$ as follows: 
$$
	\zeta_\j =
	\begin{cases}
		\zeta_{\y, (j + \lfloor\zetamin/\pi\rfloor)}, 		
		\quad
	& 	
		\text{see \eqref{eq:zetaym}, if}\quad 1\le j \le N_\streak, 
		\\
		\zetamax, \quad
	&	
		\text{otherwise,}
	\end{cases} 
$$
where $\lfloor\cdot\rfloor$ denotes the integer part. Accordingly, equations \eqref{eq:vecrdef}--\eqref{eq:ABCDdef} become: 
$$
\begin{aligned}
	\mathbf{r}_{\alpha,j}  
&\bydef 
	\veclist{\myRe\,\ISalpha(\zeta_j), \myIm\,\ISalpha(\zeta_j),\myRe\,\ITalpha(\zeta_j), \myIm\,\ITalpha(\zeta_j) }^\TT, 
\\ 
	A_{\alpha,j} &\bydef \avg{|\ISalpha|^2(\zeta_j)}  = \sigmasqalpha K_\alpha \operF{G^S_\alpha}{F_\alpha}(\zeta_j), 
\end{aligned}
$$
and similarly for $B_{\alpha,j}$, $C_{\alpha,j}$, and $D_{\alpha,j}$.
Then, introduce 
\begin{subequations}
\label{eq:rABCDj}
\begin{equation}
\label{eq:rABCDj_s}
	(\mathbf{r}_j ,  A_j ,  B_j,  C_j,  D_j, \mathbf{M}_j)_\text{s-model} 
	= \sum_{\alpha \in {\cal S}_j}
	(\mathbf{r}_{\alpha,j} ,  A_{\alpha,j} ,  B_{\alpha,j},  C_{\alpha,j},  D_{\alpha,j}, \mathbf{M}_{\alpha,j})
\end{equation}
and
\begin{equation}
\label{eq:rABCDj_t}
	(\mathbf{r}_j ,  A_j ,  B_j,  C_j,  D_j, \mathbf{M}_j)_\text{t-model} 
	= \sum_{\alpha \in {\cal T}_j}
	(\mathbf{r}_{\alpha,j} ,  A_{\alpha,j} ,  B_{\alpha,j},  C_{\alpha,j},  D_{\alpha,j}, \mathbf{M}_{\alpha,j}),
\end{equation}
\end{subequations}
where according to \eqref{eq:16Ity} and~\eqref{eq:16Ityb}, the summation sets ${\cal S}_j$ and ${\cal T}_j$ are:
\begin{equation*}
	{\cal S}_j = 
	\begin{cases} 
		\{\text{b},\text{n},s\} \! &\! \text{for}\ 1\le j \le N_\streak,
	\\
		\{\text{b},\text{n}\}\! &\!\text{for} \ j > N_\streak,
	\end{cases}
	\quad	{\cal T}_j = 
	\begin{cases} 
		\{\text{b},\text{n},t\} \!&\! \text{for}\ 1\le j \le N_\streak,
	\\
		\{\text{b},\text{n}\} \!&\!\text{for} \ j > N_\streak.
	\end{cases}
\end{equation*}
Since different image components given by~$\mathbf{r}_{\alpha,j}$ for different~$\alpha$ are assumed independent, the moments of the entire  $\mathbf{r}_j$ for a fixed $j$ and either of the two models, s-model or t-model, can be obtained by summing up the moments of the corresponding individual components given by~\eqref{eq:alphaABCD}, (\ref{eq:MABCD}): 
\begin{equation}
\label{eq:matMj}
\begin{gathered}
	\avg{\mathbf{r}_{j,\,\text{s-model}}\,\mathbf{r}_{j,\,\text{s-model}}^\TT}
	= \mathbf{M}_{j,\,\text{s-model}} 
	= \sum_{\alpha \in {\cal S}_j}
	\mathbf{M}_{\alpha,j} \,,\\
	\avg{\mathbf{r}_{j,\,\text{t-model}}\,\mathbf{r}_{j,\,\text{t-model}}^\TT}
	= \mathbf{M}_{j,\,\text{t-model}} 
	= \sum_{\alpha \in {\cal T}_j}
	\mathbf{M}_{\alpha,j} \,.
	\end{gathered}
\end{equation}
Then, the  probability density of~$\mathbf{r}_j$ for either of the two models is  given by the standard formula for multivariate Gaussian distribution: 
\begin{equation}
\label{eq:pGaussian}
	p(\mathbf{r}_j)=\frac{1}{\big( (2\pi)^4 \det \mathbf{M}_j \big)^{1/2}}
	\exp\Big({-\frac{1}{2}\mathbf{r_j}^\TT\mathbf{M}_j^{-1}\mathbf{r}_j}\Big)
	.
\end{equation}
The vector $\mathbf{r}_j$ and the matrix $\mathbf{M}_j$ in (\ref{eq:pGaussian}) must correspond to one and the same model, the s-model or t-model, see (\ref{eq:matMj}).

The overall vector~$\mathbf{R}$ combines all  vectors $\mathbf{r}$ of \eqref{eq:vecrdef} for the streak and homogeneous pairs of samples: 
\begin{equation}
\label{eq:datasetR}
	\mathbf{R} = \veclist{\mathbf{r}^\TT_1, \mathbf{r}^\TT_2, \ldots, \mathbf{r}_j^\TT, \ldots, \mathbf{r}^\TT_{N_\streak+N_\homog}}^\TT
	.
\end{equation}
As we consider each pair of samples independent, we have: %
\begin{equation}
\label{eq:logpdf}
	p(\mathbf{R}) = \prod_{j=1}^{N_\streak+N_\homog} p(\mathbf{r}_j)
	,
\end{equation}
where individual $p(\mathbf{r}_j)$ are given by~\eqref{eq:pGaussian} for either the s-model or t-model.

Let us now denote by $\mathbf{Q}$ the actual dataset vector that represents a given image. The vector $\mathbf{Q}$ has the same structure as the vector $\mathbf{R}$ of (\ref{eq:datasetR}):
\begin{equation}
\label{eq:datasetQ}
	\mathbf{Q} = \veclist{\mathbf{q}^\TT_1, \mathbf{q}^\TT_2, \ldots, \mathbf{q}_j^\TT, \ldots, \mathbf{q}^\TT_{N_\streak+N_\homog}}^\TT
	.
\end{equation}
The individual sub-vectors $\mathbf{q}_j$ in (\ref{eq:datasetQ})  correspond to the samples of the given image taken as described in Section~\ref{sec:discriminationIdea}. Each $\mathbf{q}_j$ represents one ambiguity pair and has four real-valued components arranged the same way as in (\ref{eq:vecrdef}).

The vector $\mathbf{Q}$ will provide the input for the discrimination procedure whose primary task is to tell whether it corresponds to an instantaneous or delayed target. The discrimination will be rendered by seeing whether the data $\mathbf{Q}$ fit better the s-model or the t-model, respectively. %

In addition to the input data $\mathbf{Q}$, the discrimination procedure uses the values of $\zeta_j$ that are  known. The functions  $A_j$, $B_j$, $C_j$, and~$D_j$ are  known for both the s-model and  t-model up to the factors   $\sigmasqalpha$, $\alpha \in {\cal S}_j$ or $\alpha \in {\cal T}_j$, that are not known, see (\ref{eq:ABCDdef}), (\ref{eq:rABCDj}). It is these factors that are used as optimization variables in order to achieve the best fit between the data and the model.

For a given dataset~$\mathbf{Q}$ of (\ref{eq:datasetQ}), consider the two likelihood functions \cite{Mendenhall-73} defined via \eqref{eq:matMj},
\eqref{eq:pGaussian}, and \eqref{eq:logpdf} for the two models that we have built:
\begin{subequations}
\label{eq:likelihood}
\begin{align}
\label{eq:likelihood_s}
p_\text{s-model}(\mathbf{Q}) = & \prod_{j=1}^{N_\streak+N_\homog}\frac{1}{\big((2\pi)^4 \det\mathbf{M}_{j,\,\text{s-model}} \big)^{1/2}}
	\exp\Big({-\frac{1}{2}\mathbf{q_j}^\TT\mathbf{M}_{j,\,\text{s-model}}^{-1}\mathbf{q}_j}\Big),\\
	\label{eq:likelihood_t}
	p_\text{t-model}(\mathbf{Q}) = & \prod_{j=1}^{N_\streak+N_\homog}\frac{1}{\big( (2\pi)^4 \det \mathbf{M}_{j,\,\text{t-model}} \big)^{1/2} }
	\exp\Big({-\frac{1}{2}\mathbf{q_j}^\TT\mathbf{M}_{j,\,\text{t-model}}^{-1}\mathbf{q}_j}\Big).
\end{align}
\end{subequations}
The functions $p_\text{s-model}$ and $p_\text{t-model}$ of (\ref{eq:likelihood}) depend on the unknown intensities $\sigmasqalpha$ that appear in the entries of the matrices 
 $\mathbf{M}_{j,\,\text{s-model}}$ and $\mathbf{M}_{j,\,\text{t-model}}$, see formulae \eqref{eq:ABCDdef}, \eqref{eq:MABCD}, and~\eqref{eq:matMj}.
The discrimination procedure solves two optimization problems formulated as follows: 
\begin{equation}
\label{eq:ssttOptimization}
	\breve p_s = \max_{\sigmasqz,\sigmasqn,\sigmasqs} p_\text{s-model}(\mathbf{Q})
,\quad
	\breve p_t = \max_{\sigmasqz,\sigmasqn,\sigmasqt} p_\text{t-model}(\mathbf{Q})
	,
\end{equation}
subject to $\sigmasqz,\sigmasqn,\sigmasqs,\sigmasqt \ge 0$, 
where $p_\text{s-model}(\mathbf{Q})$ and $p_\text{t-model}(\mathbf{Q})$ are defined by  (\ref{eq:likelihood_s}) and (\ref{eq:likelihood_t}), respectively. The resulting $\breve p_s$ and $\breve p_t$ yield the maximum likelihood (ML) values for the corresponding scatterer models. The classification decision, i.e., the discrimination, is made by comparing the two maxima: 
\begin{equation}
\label{eq:stAlgorithm}
\begin{array}{l}
	\text{\tt if $\breve p_t > \breve p_s$}
\\
	\text{\tt then}
\\
	\quad \text{the target  is classified as a delayed scatterer (\ref{eq:6nut}), (\ref{eq:16Ityt})} 
\\
	\text{\tt else}
\\
	\quad \text{the target  is classified as an instantaneous scatterer (\ref{eq:6nus}), (\ref{eq:16Itys}).} 
\end{array}
\end{equation} 
In other words, algorithm (\ref{eq:stAlgorithm}) attributes a given target to one of the two possible types 
based on whether $\breve p_t > \breve p_s$ or $\breve p_t < \breve p_s$. Thus, the issue of  confidence intervals  becomes important, especially in the presence of noise. Suppose, for example, that   $\breve p_t > \breve p_s$. Then, how much of a gap shall we have between $\breve p_t$ and $ \breve p_s$ to be confident that the classification of the target as a delayed scatterer is correct? This question will be addressed in the future.

\section{Performance analysis of the discrimination procedure}
\label{sec:MonteCarlo}

\begin{figure}[ht!]
\includegraphics[page=1,height=7in,clip=true]{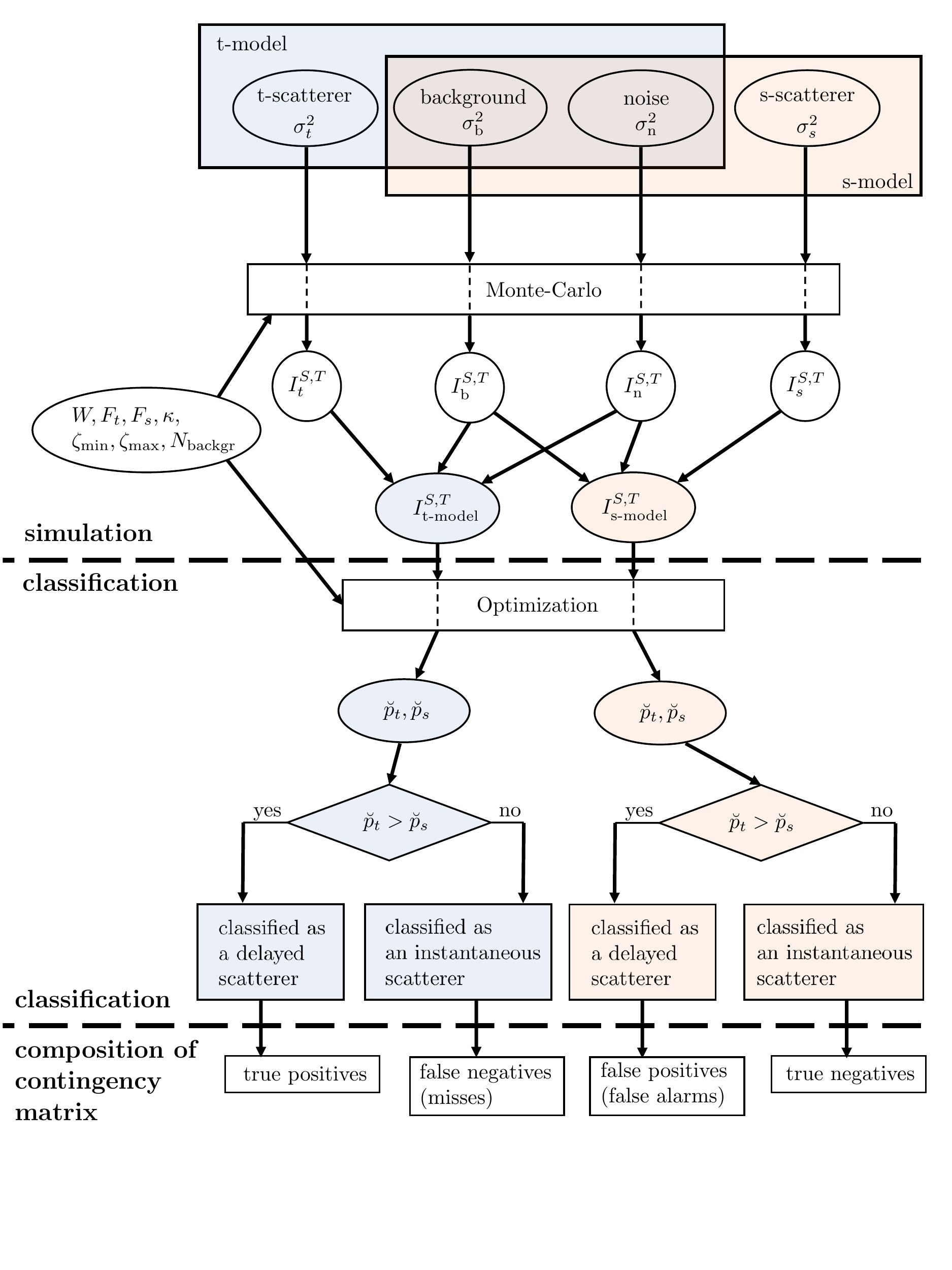}
\vspace{-1.5cm}
\caption{Performance analysis of the discrimination procedure.
} 
\label{fig:block_scheme}
\end{figure}

To assess the performance of the discrimination procedures of  	 Section~\ref{sec:st_discrimination}, we will simulate  a large number of image vectors~\eqref{eq:datasetR} using both models in~\eqref{eq:16Ity}, 
substitute the simulated vectors $\mathbf{R}$ for $\mathbf{Q}$ and thus
generate the datasets \eqref{eq:datasetQ}, 
run algorithm (\ref{eq:stAlgorithm})  on each of the datasets, and count the number of correct and incorrect classifications. As there is usually only one image available for analysis, the discrimination algorithm is not allowed to ``learn'' from the resulting statistics. 
It does not know either which of the two models has been used to obtain a given vector  \eqref{eq:datasetR}  and what the corresponding values of $\sigmasqalpha$ were.

A block diagram for performance assessment  is shown in Figure~\ref{fig:block_scheme}. 
An ensemble of sampled coordinate-delay SAR images represented by 
datasets~\eqref{eq:datasetQ} is generated using the Monte-Carlo method. We start with choosing $F_{s,t}(\zeta)$, $\zetamin$, $\zetamax$, $\kappa$, and $N_\homog$. The relative scatterer intensities, or contrasts, are defined as follows: 
\begin{equation}
\label{eq:contrastdef}
	p_\text{n} = \frac{\sigmasqn K_\text{n}}{\sigmasqz K_\text{b}}
,\quad
	q_{st} 
		= \frac{\sigmasqs K_s}{\sigmasqs K_s + \sigmasqz K_\text{b} + \sigmasqn K_\text{n}}
             	= \frac{\sigmasqt K_t}{\sigmasqt K_t + \sigmasqz K_\text{b} + \sigmasqn K_\text{n}}
	.
\end{equation}	
This allows us to calculate $\sigmasqs K_s$, $\sigmasqt K_t$, $\sigmasqz K_\text{b}$, and $\sigmasqn K_\text{n}$~accurate to a common factor. 
Note that we always take the $\sigmasqs K_s = \sigmasqt K_t $, which makes sense from the standpoint of the discrimination problem. For example, if the range-delay ambiguity is not resolved, then the statistical properties of the instantaneous and delayed images will be the same, see~\eqref{eq:3avg}. 

Each dataset  consists of the streak data and homogeneous data, as per  Section~\ref{sec:anticipatedStatistics}. To create the streak data, we generate a pair of circular Gaussian pseudo-random variables $(\ISalpha(\zetay),\ITalpha(\zetay))$ for each $\zetay$ that satisfies \eqref{eq:zetaym}  and each~$\alpha \in \{\text{b},\text{n},s,t\}$, with the moments given by~\eqref{eq:3avg}, see also %
\ref{sec:FGH}. Then we compute the sum of the resulting pairs $(\ISalpha(\zetay),\ITalpha(\zetay))$   
according to~\eqref{eq:16Itys} and~\eqref{eq:16Ityt} 
for scenarios~\eqref{eq:6nus} and~\eqref{eq:6nut}, respectively, and obtain the samples  \eqref{eq:streakST}. 
For the homogeneous data, we generate $N_\homog$ 
of pseudo-random pairs $(\IS_\text{b}(\zetamax),\IT_\text{b}(\zetamax))$ 
and $(\IS_\text{n}(\zetamax),\IT_\text{n}(\zetamax))$  
and compute the sums according to~\eqref{eq:16Ityb} 
to obtain the samples \eqref{eq:homST}. 
We  combine the streak data with the homogeneous data for each of the two target models in~\eqref{eq:16Ity} and obtain two vectors $\mathbf{R}$ of \eqref{eq:datasetR}.  The discrimination procedure  treats them as two datasets $\mathbf{Q}$ of \eqref{eq:datasetQ} ($2(N_\streak+N_\homog)$ complex numbers each). These datasets represent one  image for each of the two target models given in~\eqref{eq:6nu}. 

By repeating the foregoing procedure $\Nimg$ times, we obtain an ensemble of $2\Nimg$ ``images'' (i.e., datasets). It consists   of two sub-ensembles of $\Nimg$ images each generated using one of the two target models in~\eqref{eq:6nu}. 
After an ensemble has been generated, each image goes through the discrimination procedure, see Section~\ref{sec:st_discrimination}, and is classified as either an instantaneous (i.e., originating from the s-model) or delayed (t-model) scatterer, see~\eqref{eq:stAlgorithm}. The classification outcome  contributes to one of the four cells in table~\ref{table:contingencytable}, which is called the contingency table \cite{Mendenhall-73}. Rows of table~\ref{table:contingencytable} are determined by the actual scatterer type or, in our case, 
by the underlying model used to generate the dataset. Columns of table~\ref{table:contingencytable} are determined by the classification result. The data in table~\ref{table:contingencytable} are normalized by~$\Nimg$. 

\begin{table}[ht!]
\caption{\label{table:contingencytable}Contingency table: the entries are relative frequencies of events.}
\begin{center}
\begin{tabular} { | c | c | c |  } %
\hline
	& {\bf output: s} & {\bf output: t} %
\\ \hline
	{\bf input: s} & $1 - r_s $ & $r_s$ 	
\\ \hline
	{\bf input: t} & $r_t$ & $1 - r_t $ 
\\ \hline 
\end{tabular} 
\end{center}
\end{table}

An ideal contingency table 
would be diagonal; in other words, having 
$r_s=r_t=0$. %
If our goal is to detect delayed targets, then, e.g., $r_s$ can be identified as the false alarm ratio, see Figure~\ref{fig:block_scheme}. 
The performance of the discrimination procedure can be expressed via the values of error frequencies $r_s$ and~$r_t$. 
The quality of discrimination may depend on the scene and processing parameters, e.g., on $\zeta_{\max}$, $N_\homog$, $\kappa$, target parametrization, contrasts, sampling, etc.; some of these dependencies are demonstrated in Section~\ref{sec:ssttResults}. 

Note that the values~$r_s$ and~$r_t$ calculated using Monte-Carlo simulation are random, and hence, as metrics of the discrimination quality, contain some errors because of the finite size of the ensemble. 
Assuming $\avg{r_s}$ and $\avg{r_t}$ to be the true (and unknown) error frequencies, we find that 
each row in table~\ref{table:contingencytable} contains the averages due to the binomial distribution. Then, we have \cite{Mendenhall-73}:
\begin{equation}
\label{eq:stdpspt}
	\std(r_s) = \bigg(\frac{\avg{r_s} \cdot(1 -\avg{r_s})}{\Nimg}\bigg)^{1/2}
,\quad 
	\std(r_t) = \bigg(\frac{\avg{r_t} \cdot(1 -\avg{r_t})}{\Nimg}\bigg)^{1/2}
	,	
\end{equation}
Hence, for large~$\Nimg$ the error frequencies $\avg{r_s}$ and $\avg{r_t}$ can be approximated by their estimates $r_s$ and~$r_t$, respectively.

\section{Results of simulations}
\label{sec:ssttResults}

We have tested the discrimination between instantaneous and delayed targets using the methodology presented in Sections~\ref{sec:st_discrimination} and~\ref{sec:MonteCarlo}. Our goal was to see how the change of various problem parameters affects the quality of discrimination in percentage points defined as 
\begin{equation}
\label{eq:psptmetric}
	\text{round}\big(100\cdot (r_s+r_t)/2\big),
\end{equation}
see table~\ref{table:contingencytable}. 
With~$\Nimg = 400$, each Monte-Carlo run generates an ensemble consisting of $800$ ``images'' for calculation of the discrimination quality metrics presented in table~\ref{table:contingencytable}. 
Note that according to~\eqref{eq:stdpspt}, the stochastic errors of metric~\eqref{eq:psptmetric}  can be of the order of  $100 / \big(2\sqrt{\Nimg}\big) =  2.5$. 

The generic parameters of the simulations are: 
$\zetamin = 3\pi$, 
$\zetamax = 12\pi$ (so that $N_\streak = 10$),  
$N_\homog = 15$; 
the relative contrasts defined in~\eqref{eq:contrastdef} are  
$p_\text{n} = 0.25$ and $q_{st} = 0.4$. 
In each particular Monte-Carlo run, some of these parameters could vary. %
Our numerical simulations are as follows. 
{\renewcommand{\theenumi}{\Alph{enumi}} 
\begin{enumerate}

\newcommand{\sixMetrics}[6]{%
The corresponding quality metrics (\ref{eq:psptmetric}) were #1, #2, and~#3 for~$\kappa=0.4$, and #4, #5, and #6 for~$\kappa=1$.%
}%

\item
\label{item:vs_zetaMax}
To see how the discrimination quality depends on~$\zetamax$,  we generated three image ensembles with $\zetamax=4\pi$, $8\pi$, and~$20\pi$. 
\sixMetrics{48}{34}{6}{36}{17}{2}
Note that condition $\kappa\zetamax > \bPhi$, see~\eqref{eq:kappazetamax}, is not satisfied for the first two runs for~$\kappa=0.4$ and the first run  for~$\kappa = 1$. 
Altogether, the discrimination quality  improves with the increase of $\kappa\zetamax$. %

\item
\label{item:vs_zetaMin} To demonstrate the advantage of taking multiple $\zetaym$ (i.e., multiple delays) in the streak data, see~\eqref{eq:streakST}, we set up three ensembles with $\zetamin=3\pi$, $8\pi$, and~$12\pi$, such that 
$N_\streak = 10$, $5$, and~$1$, while $\zetamax=12\pi$. %
\sixMetrics{22}{27}{37}{11}{23}{35}
Note that for the case $N_\streak = 1$ and $\kappa \zetamax / \bPhi = 1 \cdot 12\pi/\bPhi \approx 1.6$, where condition~\eqref{eq:kappazetamax} is satisfied with a significant margin, about a third of discrimination results are still incorrect. 
We consider this case as representing the ``deterministic'' approach to discrimination (as outlined in Section~\ref{sec:ssttW}) because only a single sample of the scattering delay is taken.

\item
\label{item:vs_qalpha} 
We varied the target contrast~$q_{st}$, see~\eqref{eq:contrastdef}, by setting its values to  $0.1$, $0.3$, and $0.6$. The intensity of the background was adjusted accordingly so that the relative intensity of noise  always remained 0.1, i.e., 
$p_\text{n} = 0.1 / (0.9-q_{st})$. 
\sixMetrics{47}{35}{17}{43}{29}{5}
Obviously, discrimination of low-contrast targets is less reliable than that of high-contrast targets.

\item
\label{item:vs_numBackground} 
The number of homogeneous pairs of samples \eqref{eq:homST} did not noticeably affect the quality of discrimination. We have tried various combinations of parameters, including $N_\homog = 0$, and the  effect never exceeded 5\%.

\end{enumerate}
} %

\section{Discussion}
\label{sec:ssttDiscussion}

We have demonstrated a functioning methodology of distinguishing between delayed and instantaneous scatterers in coordinate-delay SAR images. %
To make this approach more practical, several issues still need to be addressed. %

\begin{itemize}

\item 
Testing the discrimination of targets against an 
inhomogeneous or textured background~\cite{oliver-98} 
and with delayed and inhomogeneous targets having non-singular support of the reflectivity function (cf.\ \eqref{eq:modeltz} and~\eqref{eq:modelsz}).

\item
Fully taking into account the correlation between image samples rather than restricting it to  (S,T)-pairs as explained in Section~\ref{sec:anticipatedStatistics}. This opens a venue to increasing the number of samples of each image. 

\item
Obtaining the confidence level for a given discrimination. 

\item 
Automatic selection of the reference target position~$\zd$. The coordinates $\zdO,\zdT$ may be included into the set of optimization variables in~\eqref{eq:ssttOptimization}. 

\item
Testing the discrimination method on a broader set of functions~$F_t(\zeta)$ and~$F_s(\zeta)$. In particular, these function may correspond to certain physical mechanisms of delayed scattering, including the cases where the actual returned signal has a deterministic component. Examples of such mechanisms can be the Foldy-Lax dispersion model, see \cite[Section~9.9]{devaney-12}, or the waveguide model of~\cite{borden-98}. 

\end{itemize}

Note that the discrimination procedure  assumes a certain form of the functions $F_t(\zeta)$ in \eqref{eq:mutmut} and~$F_s(\zeta)$ in \eqref{eq:musmus}, see, e.g., (\ref{eq:Heaviside}). For testing, we generated the data that matched 
the assumptions built into the discrimination algorithm (see Section~\ref{sec:MonteCarlo}). In practice, this may be the case 
 if we have some a priori knowledge about the possible target.

 However, in real applications we may have to deal with the targets characterized by $F_t(\zeta)$ or~$F_s(\zeta)$ that are not known to the remote system. If these functions differ significantly  from those used in the discrimination procedure, then the optimization problems~\eqref{eq:ssttOptimization} lose their relevance. To remedy this situation, one can represent $F_{s,t}(\zeta)$ as expansions with respect to a specially chosen basis $\{F_i(\zeta)\}$. 
As the functionals~$G^{S,T}_\alpha$ and~$H_\alpha$ are linear, see \ref{sec:FGH}, equations~\eqref{eq:3avg} will change according to
\begin{equation}
\label{eq:alpham}
\begin{aligned}
	\sigmasqalpha \operF{G^{S,T}_\alpha}{F_\alpha}(\zeta) 
\mapsto&\> 
	\sum_{i=1}^M \sigma^2_{\alpha,i}\operF{G^{S,T}_\alpha}{F_i}(\zeta)
	,
\\
	\sigmasqalpha \operF{H_\alpha}{F_\alpha}(\zeta) 
\mapsto&\> 
	\sum_{i=1}^M \sigma^2_{\alpha,i}\operF{H_\alpha}{F_i}(\zeta)
	,
\end{aligned}
\end{equation}
where the basis functions~$F_i(\zeta)$ are known. Then, each unknown~$\sigmasqalpha$ in the optimization problems will be replaced with~$M$ expansion coefficients, $\sigma^2_{\alpha,i}$, while the rest of the procedure remains the same.

The choice of the basis $\{F_i(\zeta)\}$ and its dimension $M$ is  central for 
the  approach~\eqref{eq:alpham}. The case we considered in this paper is $M=1$ and $F_1(\zeta)$ given by~\eqref{eq:Heaviside}. 
By increasing~$M$ and properly selecting~$F_i$, we can fit the unknown functions~$F_s$ and~$F_t$ better. As, however, indicated in Section~\ref{sec:discriminationIdea}, the  sampling rate w.r.t.\ $\zetay$ in the streak data is bounded from below, see~\eqref{eq:zetaym}. Hence, there are only finitely many samples $F_s(\zetaym)$ and~$F_t(\zetaym)$, and choosing $M \ge N_\streak $ %
will imply overfitting. 
We leave this and related topics (e.g., whether we should require $\sigmasq_{\alpha,i} \ge 0$ for all $\alpha$ and~$i$) for a future study.

An even more serious modification of the current approach may be required to  accommodate the wide-angle and full circular apertures, where the reflectivity in~\eqref{eq:linscattering} can no longer be considered independent of~$\x$, or delays smaller than~$1/B$ that make the streak shorter than one pixel in range, see Figure~\ref{fig:sstt_streak}. 

\section*{Acknowledgements}

We would like to thank Profs.\ M.~Cheney (Colorado State University) and A.~Doerry (University of New Mexico; Sandia) for fruitful discussions. 
This material is based upon work supported by the US Air Force Office of Scientific Research (AFOSR) under awards number FA9550-14-1-0218 and  FA9550-17-1-0230.
The first author also acknowledges support 
by the National Science Foundation under Grant No.~DMS-1439786 while he was in residence at the Institute for Computational and Experimental Research in Mathematics (Brown University, Providence, RI) during the Fall Semester of 2017.

\appendix

\section{Second moments of image components}
\label{sec:FGH}

Here we calculate the right hand sides of~\eqref{eq:3avg} for different scatter types denoted by~$\alpha$. 
We need to recall formula \eqref{eq:Iconvconv} for the image:
\begin{equation}
\label{eq:app:Iconvconv}
	I(t_\y,\y) 
	= \int_0^\infty dt_\z \int \, d\z\, \nu(t_\z,\z) 
	W(t_\y,\y;t_\z,\z)
	,
\end{equation}
and expression~\eqref{eq:etazetapsiW} for the imaging kernel 
\begin{equation}
\label{eq:app:WviaPhi}
	W(t_\y,\y;t_\z,\z) 
	= 
	N\tau \exp\Big({-2i\frac{\omega_0}{B}\zeta}\Big) \cdot \Phi\big(\eta, \kappa (\zeta - B(t_\y-t_\z)/2) \big)  \cdot \sinc \zeta
	,
\end{equation}
where
\begin{equation*}
\begin{aligned}
	\kappa =  \phiT^2 \frac{\omega_0}{B}
,\quad 
	\eta 
=%
	\kOtheta \phiT(y_1-z_1)
,\quad %
	\zeta
=%
	\frac{B}{\omega_0}\kOtheta(y_2 - z_2) + B \frac{t_\y-t_\z}{2}
	,
\end{aligned}
\end{equation*}
see \eqref{eq:kappadef} and~\eqref{eq:etazetapsi}.
In what follows, we will use these expressions with different formulations for~$\nu(t_\z,\z)$. 

To simplify calculations, functions $F_t(\zeta)$ and $F_s(\zeta)$ in this appendix are taken as 
$F_t(\zeta) = F_s(\zeta) = (1+\sign \zeta)/2$ rather than the indicator function~\eqref{eq:Heaviside}, while $\zetamax$ in Sections \ref{sec:discriminationIdea} and~\ref{sec:ssttResults} acts as a parameter of sampling. 
Similarly to Section~\ref{sec:imagesInhom} (see discussion around \eqref{eq:IsIt} and~\eqref{eq:IsIt2}),
as long as we take $\zetamax \gg \pi$, which is true for all cases considered in Section~\ref{sec:ssttResults}, 
the effect on the 
values of integrals in this appendix 
is insignificant.

\subsection*{Homogeneous scatterer~$\nu_\text{{\rm b}}(t_\z,\z)$ in  \eqref{eq:modelz}--\eqref{eq:muzmuz}:}
\begin{equation}
\label{eq:app:modelz}
	\nu(t_\z,\z) = 
	\nuz(t_\z,\z) = 
	\delta(t_\z)\muz(\z)
,\quad 
	\big\langle\overline{\muz(\z_\text{a})} \muz(\z_\text{b}) \big\rangle 
=
	\sigmasqz \delta (\z_\text{a} - \z_\text{b})
	.
\end{equation}
Substituting~\eqref{eq:app:modelz} into \eqref{eq:app:Iconvconv}--\eqref{eq:app:WviaPhi}, we obtain
\begin{equation}
\label{eq:app:Iz}
\begin{aligned}
	I_\text{b}(t_\y,\y) &= N\tau \int \muz(\z) \exp\Big({-2i\frac{\omega_0}{B} \zeta_0}\Big) \sinc \zeta_0\; \Phi(\eta,\kappa\zeta_0)\,d\z,
\\ & \qquad
\text{where}\quad 
	\zeta_0 = \zeta\Big|_{t_\z=0}.
\end{aligned}
\end{equation}
Replacing~$t_\y$ with a dimensionless argument
\begin{equation}
\label{eq:app:zetaydef}
	\zetay = Bt_\y/2, 
\end{equation}
using the delta functions in~\eqref{eq:app:modelz}, and performing a change of integration variables, we can obtain 
\begin{equation}
\label{eq:intzdeeta}
	\avg{|I_\text{b}(\zetay,\y)|^2} 
	= \sigmasqz \cdot N^2\tau^2 \frac{\omega_0}{B\kOtheta} \frac{1}{\kOtheta\phiT} \cdot 
	\int d\zeta_0 \sinc^2 \zeta_0 \int d\eta\; \big|\Phi\big(\eta,\kappa(\zeta_0-\zetay)\big)\big|^2
\end{equation}
Obviously, the expression on the right hand side does not depend of~$\y$. A less expected result is that it  does not depend on~$t_\y$ (or~$\zetay$) either. In order to prove the latter statement, we notice that expression~\eqref{eq:Phidef} can be formally considered as a Fourier transform $s \leftrightarrow v_1$:
$$
\begin{aligned}
	\Phi(v_1,v_2) &= 
	\int_{-1/2}^{1/2} \exp({2iv_1s})\exp({iv_2s^2})\,ds = 
	{\cal F}\big[\Psi_{v_2}(s)\big](2v_1),
\\ & \qquad
	\text{where}\quad 
	\Psi_{v_2}(s) = \exp({iv_2s^2})\chi_1(s)	
	,
\end{aligned}
$$
and the indicator function~$\chi_1$ is defined similarly to $\chi_\tau$ in~\eqref{eq:indicator}. Then, due to the Parseval's theorem, we have the following identity: 
\begin{equation}
\label{eq:ParsevalPhi}
\begin{aligned}
	\int \Phi(\eta,a) \overline{\Phi(\eta,b)}\,d\eta 
=&\>
	 \pi \int \Psi_a(s) \overline{\Psi_b(s)}\, ds 
\\ =&\> 
	 \pi\int_{-1/2}^{1/2} \exp({i(a-b)s^2})\,ds = \pi\Phi(0,a-b)
	 .
\end{aligned}
\end{equation}
Formula~\eqref{eq:ParsevalPhi} immediately evaluates the interior integral in~\eqref{eq:intzdeeta} to~$\pi$, and then the outer integral yields another~$\pi$. Hence, formula~\eqref{eq:intzdeeta} reduces to 
\begin{equation}
\label{eq:app:Kz}
	\avg{|I_\text{b}|^2} 
	= \sigmasqz K_\text{b} 
,\quad\text{where}\quad
	K_\text{b}  = 
	N^2\tau^2 \frac{\omega_0}{B\kOtheta} \frac{1}{\kOtheta\phiT} \cdot 
	\pi^2
	.
\end{equation}

Using definitions~\eqref{eq:streakST} for~$\IS$ and~$\IT$ and 
taking an arbitrary~$\zd$, we write the following expression for the cross-correlation:
\begin{multline}
\label{eq:intzSTdeeta}
	\avg{\ISz(\zetay) \roverline{\ITz}(\zetay)} 
	= \sigmasqz \cdot N^2\tau^2 \frac{\omega_0}{B\kOtheta} \frac{1}{\kOtheta\phiT} \cdot 
	\int d\zeta_0 \sinc^2 \zeta_0 
\\ \cdot 
	\int d\eta\; 
	\Phi(\eta,\kappa\zeta_0)
	\overline{	\Phi\big(\eta,\kappa(\zeta_0-\zetay)\big)} 
	.
\end{multline}
Applying once again formula~\eqref{eq:ParsevalPhi} to the interior integral in~\eqref{eq:intzSTdeeta}, we obtain
$\pi \Phi(0,\kappa\zetay)$, and, likewise, the outer integration can then be performed. Altogether, from~\eqref{eq:intzdeeta} and~\eqref{eq:intzSTdeeta}, in notations of~\eqref{eq:3avg} and taking into account~\eqref{eq:app:Kz}, we write 
\begin{equation}
\label{eq:app:FGz}
	\operF{G^S_\text{b}}{F_\text{b}}(\zeta) = \operF{G^T_\text{b}}{F_\text{b}}(\zeta) = 1	
,\quad
	\operF{H_\text{b}}{F_\text{b}}(\zeta) = \Phi(0,\kappa\zeta)
	, 
\end{equation}
where $F_\text{b}$ may be formally defined as $F_\text{b}(\zeta) \equiv 1$. 
Relations \eqref{eq:app:Kz} and~\eqref{eq:app:FGz} define the right hand sides of~\eqref{eq:3avg} for $\alpha=\text{b}$, i.e., for the image component representing the instantaneous homogeneous background.  

\subsection*{Delayed point scatterer (t-scatterer) $\nut(t_\z,\z)$ in \eqref{eq:modeltz}--\eqref{eq:mutmut}:}
\begin{equation}
\label{eq:app:modeltz}
	\nu(t_\z,\z) = 
	\nut(t_\z,\z) 
	=  
	\mut(t_\z)\delta(\z-\zd)
,\quad
	\big\langle\overline{\mut(t_\text{a})} \mut(t_\text{b}) \big\rangle 
	= 
	\sigmasqt F_t(Bt_\text{a}/2) \delta (t_\text{a} - t_\text{b}) 
	.
\end{equation} 
Substituting~\eqref{eq:app:modeltz} into \eqref{eq:app:Iconvconv}--\eqref{eq:app:WviaPhi}, we obtain
\begin{equation}
\label{eq:app:It}
	I_t(\zetay,\etad,\xid) = N\tau 
	\cdot \Phi\big(\etad,\kappa\xid\big)
	\int_0^\infty \mut(t_\z) \exp\Big({-2i\frac{\omega_0}{B} \zetad}\Big) \sinc \zetad 
	\,dt_\z
\end{equation}
where
\begin{equation}
\label{eq:etadxid}
	\etad = \kOtheta\phiT(y_1 - \zdO)
,\quad
	\xid = \frac{B\kOtheta}{\omega_0}(y_2-\zdT)
,\quad 
	\zetad = \xid + \zetay - \frac{Bt_\z}{2}
	,
\end{equation}
and $\zetay = Bt_\y/2$ as in~\eqref{eq:app:zetaydef}. Note that unlike~\eqref{eq:etazetapsi}, dimensionless arguments of~$I_t$ in~\eqref{eq:app:It} are not aligned with the ambiguity lines, which can be expressed as $(\etad = \text{const}, \; 
\xid + \zetay = \text{const})$. 

From \eqref{eq:app:modeltz} and~\eqref{eq:app:It},   we can derive the following relations for expectations of the intensity and correlation along the ambiguity line:
\begin{equation}
\label{eq:app:3It}
\begin{aligned}
	\avg{|I_t(\zetay,\etad,\xid)|^2}
&= 
	\sigmasqt	\cdot N^2\tau^2\frac{2}{B} 
	\breve{F}_t(\zetay+\xid) 
	\cdot |\Phi (\etad, \kappa\xid)|^2
	,
\\
	\avg{\roverline{I_t(\zetay,\etad,\xid)} I_t(0,\etad,\xid + \zetay)} 
&=
	\sigmasqt	\cdot N^2\tau^2\frac{2}{B} 
	\breve{F}_t(\zetay+\xid)		
\\ & \quad 	
	\cdot 
	\Phi \big(\etad, \kappa(\xid+\zetay)\big)
	\overline{	\Phi(\etad, \kappa\xid) } 
	,
\end{aligned} 
\end{equation}
where  
\begin{equation}
\label{eq:brevedef}
	\breve{F}_t(\zeta) \bydef \int_0^\infty F_t(\zeta') \sinc^2(\zeta-\zeta') \,d\zeta'
	.
\end{equation}
If  we take $F_t(\zeta) = (1+\sign(\zeta))/2$ as in~\eqref{eq:Heaviside}, or any other function~$F_t$ that slowly varies on the interval of~$(\zeta-\pi,\zeta+\pi)$ for $|\zeta| \gtrsim \pi$, 
then $\breve{F}_t(\zeta) \approx \pi F_t(\zeta)$ for $|\zeta| \gtrsim \pi$. Hence, the arguments of $\breve{F}_t$ in~\eqref{eq:app:3It} can be understood as follows: 
scatterer~\eqref{eq:app:modeltz} affects the image as long as $(\zetay+\xid) \gtrsim \pi$; in other words, the ambiguity line (see Figure~\ref{fig:sstt_streak}) drawn through $(t_\y,\y)$ should intersect the ray $\{(t'_\y,\y') \; | \; \y'=\zd;\; Bt'_\y/2>\pi\}$. At the same time, for $ (\zetay+\xid) \lesssim -\pi$, the scatterer has no effect on the image, and the transition area width is of the order of range resolution. 

With the help of the anti-derivative 
$$
	\sinc^2\zeta = \Big(\text{Si}(2\zeta)-\sin\zeta \sinc \zeta \Big)',
$$
where $\text{Si}(\zeta) = \int_0^\zeta \sinc \zeta' \, d\zeta' $~is the sine integral, we can calculate 
$$
	\breve{F}_t(\zeta) = \frac{\pi}{2} + \text{Si}(2\zeta)-\sin\zeta \sinc \zeta
$$ 
for the simple case when $F_t(\zeta) $ is the Heaviside function used in Section~\ref{sec:ssttResults}. 
In order to normalize expressions~\eqref{eq:app:3It}, we take $\zeta\gg\pi$ in~\eqref{eq:brevedef}; this yields $\breve{F}_t(\zeta) \approx \pi$. 
For the the right hand sides of \eqref{eq:streakISTalpha_s}--\eqref{eq:streakISTalpha_t}, 
we are only interested in $\etad = 0$ and~$\xid = 0$.  Hence, for~$\alpha=t$ in~\eqref{eq:3avg}, we have the following expressions: 
\begin{equation*}
\begin{split}
	K_t = N^2\tau^2\frac{2}{B}\pi
,\quad
	\operF{G^S_t}{F_t}(\zeta) = \frac{1}{\pi} |\Phi (0, \kappa\zeta)|^2 \breve{F}_t(\zeta),
\\ 
	\operF{G^T_t}{F_t}(\zeta) = \frac{1}{\pi} \breve{F}_t(\zeta)
,\quad
	\operF{H_t}{F_t}(\zeta) = \frac{1}{\pi} \Phi(0,\kappa\zeta) \breve{F}_t(\zeta)
	.
\end{split}
\end{equation*}

\subsection*{Inhomogeneous instantaneous scatterer (s-scatterer) $\nus(t_\z,\z)$ in \eqref{eq:modelsz}--\eqref{eq:musmus}:}
\begin{equation}
\label{eq:app:modelsz}
\begin{split}
	\nu(t_\z,\z) = 
	\nus(t_\z,\z) = 
	\delta(t_\z) \delta(z_1-\zdO) \mus(z_2-\zdT),
\\
	\big\langle\overline{\mus(s_\text{a})} \mus(s_\text{b}) \big\rangle
=
	\sigmasqs F_s(B\kOtheta s_\text{a} / \omega_0) \delta (s_\text{a} - s_\text{b}) 	
	.
\end{split}
\end{equation}
Substituting~\eqref{eq:app:modelsz} into \eqref{eq:app:Iconvconv}--\eqref{eq:app:WviaPhi} and using the notation $s=z_2-\zdT$, we obtain
\begin{equation*}
	I_s(\zetay,\etad,\xid) = N\tau \int_0^\infty \mus(s) \exp\Big({-2i\frac{\omega_0}{B} \zeta_0}\Big) \sinc \zeta_0\; 
	\Phi\big(\etad,\kappa(\xid-\xis) \big)\,ds
	,
\end{equation*}
where 
$$
	\xis = \frac{B}{\omega_0}\kOtheta s
,\quad
	\zeta_0 = \frac{B}{\omega_0}\kOtheta(y_2-z_2)+B\frac{t_y}{2} 
	= \xid + \zetay - \xis 
	,
$$
cf.\ \eqref{eq:app:Iz} and~\eqref{eq:etadxid}. 
Similarly to~\eqref{eq:app:3It}, we obtain for the s-scatterer the following relations: 
\begin{multline}
\label{eq:app:3Is}
	\avg{|I_s(\zetay,\etad,\xid)|^2}
=
	\sigmasqs	\cdot N^2\tau^2\frac{\omega_0}{B\kOtheta} 
\\  \shoveright{ \cdot 
	\int_0^\infty
	\big|\Phi \big(\etad, \kappa(\xid - \xis)\big) \big|^2
	\sinc^2(\xid + \zetay - \xis) F_s(\xis)\,d\xis
	,
\quad }
\\ \shoveleft{
	\avg{\roverline{I_s(\zetay,\etad,\xid)} I_s(0,\etad,\xid + \zetay)} 
=
	\sigmasqs	\cdot N^2\tau^2\frac{\omega_0}{B\kOtheta} 
}
\\  \shoveright{ \cdot 
	\int_0^\infty
	\Phi \big(\etad, \kappa(\xid + \zetay - \xis)\big) 
	\overline{\Phi \big(\etad, \kappa(\xid - \xis)\big)}
	\sinc^2(\xid + \zetay - \xis) F_s(\xis)\,d\xis
	.
}
\end{multline}
The argument of~$\sinc^2$ on the right sides of~\eqref{eq:app:3Is} implies that the inhomogeneous scatterer~\eqref{eq:app:modelsz} affects the image as long as $(\xid + \zetay) \gtrsim \pi$ (cf.~\eqref{eq:app:3It}).

In order to achieve a proper normalization of~$G^{S,T}_s$, we use the same approach as for~\eqref{eq:app:3It}. Namely, 
consider the upper formula in~\eqref{eq:app:3Is} for $\zetay=0 $  and $\etad = 0$ when $F_s$ is the Heaviside function~\eqref{eq:Heaviside}. Shifting the integration variable, we reduce this formula to 
\begin{equation}
\label{eq:downrangeIs}
	\avg{|I_s(0,0,\xid)|^2}
=  
	\sigmasqs	\cdot N^2\tau^2\frac{\omega_0}{B\kOtheta} 
	\int_{-\xid}^\infty
	\big|\Phi \big(0, \kappa\zeta\big) \big|^2 
	\sinc^2 \zeta  \,d\zeta
	,
\end{equation}
where the dependence of the right hand side on~$y_2$ is via~$\xid$, see~\eqref{eq:etadxid}. 
When $\xid \gg 1$ (or, equivalently, $|y_2-\zdT| \gg \Delta_\rng$), 
the integral on the right hand side does not depend on~$\xid$, which is expected if we realize that the left hand side of~\eqref{eq:downrangeIs} is the standard SAR image taken downrange by many resolution sizes with respect to the inhomogeneity at~$\z=\zd$ due to the the s-target. We normalize this integral by its value at $\kappa = 0$, hence, 
$$
	K_s = N^2\tau^2\frac{\omega_0}{B\kOtheta} \pi 
	.
$$
Taking $\etad=0$ and $\xid = 0$ similarly to the case of t-scatterer, we obtain 
\begin{equation*}
\begin{aligned}
	\operF{G^S_s}{F_s}(\zeta) 
&=\> 
	\frac{1}{\pi}
	\int_0^\infty
	\big|\Phi \big(0, \kappa(\zeta - \xi)\big) \big|^2 
	\sinc^2(\zeta - \xi) F_s(\xi)\,d\xi
	,
\\
	\operF{G^T_s}{F_s}(\zeta)
&=\> 
	\frac{1}{\pi}
	\int_0^\infty
	\big|\Phi \big(0, - \kappa\xi \big) \big|^2
	\sinc^2(\zeta - \xi) F_s(\xi)\,d\xi
	,
\\
	\operF{H_s}{F_s}(\zeta)
&=\> 
	\frac{1}{\pi}
	\int_0^\infty
	\Phi \big(0, \kappa(\zeta - \xi)\big)
	\overline{\Phi \big(0, - \kappa\xi \big)}
	\sinc^2(\zeta - \xi) F_s(\xi)\,d\xi
	.
\end{aligned} 
\end{equation*}
These expressions are used in~\eqref{eq:3avg} for~$\alpha=s$. 

\subsection*{Terms $I_\text{{\rm n}}$ in \eqref{eq:16Ity} and~\eqref{eq:16Ityb}}

We have chosen the form of the noise term in \eqref{eq:16Ity} and~\eqref{eq:16Ityb} as $I_\text{n}(t,\y)$ to achieve uniformity of notations for the image components in~\eqref{eq:3avg} and on. In the absence of any specific information about the properties of the noise, we choose to define it as  an uncorrelated additive term in~\eqref{eq:streakISTalpha_s}--\eqref{eq:streakISTalpha_t}, so there is no underlying reflectivity function $\nu(t,\z)$ in 
\eqref{eq:app:Iconvconv} for~$I_\text{n}$. In the notations of~\eqref{eq:3avg}, we can formally set   
$$
	\operF{G^{S,T}_\text{n}}{F_\text{n}}(\cdot) \equiv 1
,\quad	
	\operF{H_\text{n}}{F_\text{n}}(\cdot) \equiv 0
,\quad	
	K_\text{n}=1
	, 
$$
but the only essential part is choosing the noise level: it is defined relative to the reflectivity of the homogeneous background via the constant~$p_\text{n}$, see~\eqref{eq:contrastdef}. So, in order to implement the   noise term in equations \eqref{eq:16Ity}, we generate both $\IS_\text{n}$ and~$\IT_\text{n}$ for \eqref{eq:streakST} and~\eqref{eq:homST} as uncorrelated pseudo-random circular Gaussian (as in~\eqref{eq:muzasasum},\eqref{eq:muzReIm}) numbers with the variance given by~$p_\text{n}$.


\end{document}